\documentclass[conference]{IEEEtran}
\IEEEoverridecommandlockouts
\usepackage{cite}
\usepackage{amsmath,amssymb,amsfonts}
\usepackage{algorithmic}

\usepackage{textcomp}
\usepackage{xcolor}
\usepackage{enumitem}
\usepackage{booktabs} 
\usepackage{stfloats}
\usepackage{graphicx} 
\usepackage{subcaption} 

\usepackage[ruled,vlined]{algorithm2e}

\usepackage{url}
\usepackage[T1]{fontenc}

\def\BibTeX{{\rm B\kern-.05em{\sc i\kern-.025em b}\kern-.08em
    T\kern-.1667em\lower.7ex\hbox{E}\kern-.125emX}}
\begin{document}

\title{A Two-Stage ISAC Framework for Low-Altitude Economy Based on 5G NR Signals
}

\author{\IEEEauthorblockN{Haisu Wu, Hong Ren,~\IEEEmembership{Member,~IEEE},  Cunhua Pan,~\IEEEmembership{Senior Member,~IEEE}, Boshi Wang, Jun Tang, \\Haoyang Weng, Feng Shu,~\IEEEmembership{Member,~IEEE}, Jiangzhou Wang,~\IEEEmembership{Fellow,~IEEE}}
\thanks{The work of Cunhua Pan was supported in part by the Jiangsu Outstanding Youth Fund BK20240071 and National Natural Science Foundation of China under Grant 62350710796. This work of Hong Ren was supported in part by National Natural Science Foundation of China under Grant No. 62471138.}
\thanks{H. Wu, H. Ren,  C. Pan, B. Wang, J. Tang, H. Weng  and J. Wang are with National Mobile Communications Research Laboratory, Southeast University, Nanjing, China. (e-mail: {wuhaisu, hren, cpan, $\text{boshi\_wang}$, 220241016, 220241245, j.z.wang}@seu.edu.cn).}
\thanks{Feng Shu are with the School of
Information and Communication Engineering, Hainan University, Haikou
570228, China (email:shufeng@hainanu.edu.cn).}
\thanks{\itshape{Corresponding authors:}  Hong Ren and Cunhua Pan.}
}

\maketitle

\begin{abstract}
The evolution of next-generation wireless networks has spurred the vigorous development of the low-altitude economy (LAE). To support this emerging field while remaining compatible with existing network architectures, integrated sensing and communication (ISAC) based on 5G New Radio (NR) signals is regarded as a promising solution. However, merely leveraging standard 5G NR  signals, such as the Synchronization Signal Block (SSB), presents fundamental limitations in sensing resolution. To address the issue, this paper proposes a  two-stage coarse-to-fine sensing framework that utilizes standard 5G NR initial access signals augmented by a custom-designed sparse pilot structure (SPS) for high-precision unmanned
aerial vehicles (UAV) sensing. In Stage I, we first fuse information from the SSB, Type\#0-PDCCH, and system information block 1 (SIB1) to ensure the initial target detection. In Stage II, a refined estimation algorithm is introduced to overcome the resolution limitations of these signals. Inspired by the sparse array theory, this stage employs a novel SPS,  which is inserted into  resource blocks (RBs) within the CORSET\#0 bandwidth. To accurately extract the off-grid range and velocity parameters from these sparse pilots, we develop a corresponding high-resolution algorithm based on the weighted unwrapped phase (WUP) technique and the RELAX-based iterative method.  Finally, the density-based spatial clustering of applications with noise (DBSCAN) algorithm is adopted to prune the redundant detections arising from beam overlap.  Comprehensive simulation results demonstrate the superior estimation accuracy and computational efficiency of the proposed framework in comparison to other techniques.

\end{abstract}

\begin{IEEEkeywords}
Integrated sensing and communication (ISAC), 5G New Radio (5G NR), multiple-input multiple-output orthogonal frequency division multiplexing (MIMO-OFDM),  low-altitude economy (LAE)
\end{IEEEkeywords}

\section{Introduction}

\IEEEPARstart{T}{he} evolution of future sixth-generation (6G)  wireless networks is driven by the demands of emerging applications, with the low-altitude economy (LAE) at the forefront\cite{jiang2025integrated}. As a new economic form, the LAE utilizes low-altitude airspace to conduct diverse flying activities, which promises to deliver revolutionary productivity advancements in sectors such as logistics delivery, environmental monitoring, and intelligent inspection\cite{wang2025toward,song2025trustworthy}. As this large-scale economic activity is expected to involve millions of aircraft\cite{gupta2015survey}, the supporting network needs to provide a dual capability: massive, real-time data exchange for communication and continuous, high-precision sensing for safe and uniform regulation\cite{hoque2021iotaas,wu2021comprehensive}.

To address these converging requirements, integrated sensing and communication (ISAC) has emerged as a critical enabling technology\cite{liu2020joint, liu2022integrated}. By multiplexing 5G-Advanced (5G-A) or future 6G communication signals, ISAC can simultaneously support high-speed data transmission for unmanned aerial vehicles (UAVs) and provide effective flight supervision. Specifically, the seamless coverage provided by the communication network is leveraged for the effective perception, localization, and surveillance of UAVs. This capability, in turn, assists in UAV navigation, path planning, and airspace management, which is essential for the prosperity of the LAE\cite{gupta2015survey, jiang2025integrated}.

Building on this motivation, considerable research has focused on the sensing of UAVs utilizing   terrestrial base stations (BSs) to support the LAE. In  monostatic scenarios, for example, the authors of \cite{luo2024integrated} proposed a practical ISAC framework for cluttered environments and addressed the power allocation between communication and sensing beams. In addition, the authors of \cite{pucci2022system} introduced an efficient algorithm for jointly  angle, velocity, and range estimation in the OFDM-based ISAC system, which includes a pruning algorithm for redundant target points. In multistatic scenarios, \cite{tang2025cooperative} investigated the UAV parameter estimation by employing the tensor decomposition  technique and proposed a minimum spanning tree (MST)-based multi-BS data association scheme. Moreover, an extended Kalman filter (EKF) framework was introduced in  [4]  to predict the UAV's movement and developed a corresponding handover strategy.

Notably, the aforementioned research often operates without considering the practical constraints of standardized protocols like 5G NR\cite{3gpp_38_211,3gpp_38_212, 3gpp_38_213,3gpp_38_214} or Wi-Fi \cite{ieee80211ax}, focusing instead on theoretical aspects such as beamforming or parameter extraction. To foster practical deployment, both academia and industry have begun to focus on implementing sensing within these protocol frameworks, and the investigations generally fall into two categories: data-based and pilot-based sensing.

In  monostatic data-based sensing scenarios,  BS’s \textit{ priori} knowledge of transmitted data can be used to estimate UAV parameter estimation via techniques such as matched filtering and data cancellation, while beamforming can steer energy toward potential targets\cite{keskin2025fundamental}. This principle can be applied within the 5G NR standard, where several downlink channels serve as potential signals of opportunity for wireless sensing. Early studies have explored the feasibility of using data-centric channels, such as the Physical Downlink Control Channel (PDCCH), the Physical Downlink Shared Channel (PDSCH), and the associated Demodulation Reference Signals (DMRS)\cite{keskin2025fundamental}. For instance, the theoretical ambiguity functions of PDCCH and PDSCH were analyzed in [58]. However, such approaches are fundamentally dependent on active downlink data transmissions, rendering the sensing capability opportunistic rather than persistent. In addition, leveraging user-centric data channels like PDSCH for sensing poses significant privacy risks, as sensitive user information could be exposed to malicious eavesdroppers.

To overcome the reliance on user data and the associated privacy concerns, research has shifted towards dedicated reference signals that are independent of communication traffic. The Positioning Reference Signal (PRS) and the Channel State Information Reference Signal (CSI-RS) are potential candidates, as they are specifically designed for positioning and channel estimation. The work in \cite{wei20225g} investigated the application of the PRS for radar sensing by proposing a multi-frame estimation scheme and deriving its Cramér-Rao lower bounds. In addition, \cite{golzadeh2024joint} introduced a novel irregular resource pattern to suppress the inherent range-velocity ambiguities of the comb-like PRS structure. Despite the advantages, the dedicated signals like PRS and CSI-RS are typically transmitted only upon specific network configuration and are not always-on in most operational networks, which limits their ubiquitous applicability for persistent sensing.

In contrast, the Synchronization Signal and PBCH Block (SSB) is a fundamental component of the 5G NR frame structure that is broadcast periodically, which enables it an ideal and highly reliable signal of opportunity. Consequently, SSB-based sensing has been extensively investigated\cite{yang2023angle,abratkiewicz2023ssb,awad2025ssb,chen2021carrier,liu2023machine,yang2024dynamic}. For instance, the authors of \cite{yang2023angle} leveraged the beam-sweeping nature of the SSB, utilizing RSRP measurements at the user equipment (UE) to achieve spatial environment mapping. Similarly, \cite{abratkiewicz2023ssb} exploited the periodicity of the SSB for passive coherent location (PCL). However, the study also  highlighted that the maximum unambiguous velocity of PCL is severely constrained by the SSB's long transmission periodicity (e.g., 10 ms minimum). While the study in \cite{awad2025ssb} addressed UAV detection using the SSB, its performance was limited: angular resolution is constrained by the wide beamwidth of the SSB, and velocity estimation accuracy is limited by the short duration of the per-beam OFDM transmission. To further address these limitations, the work in \cite{golzadeh2023downlink} proposed to combine the SSB with downlink control information (DCI) and system information block 1 (SIB1) signals to mitigate the radar ambiguity problem. Notably, while the 5G NR standard allocates a bandwidth part of $N_{\text{RB}}^\text{DL,BWP}$ resource blocks (RBs) for PDSCH carrying SIB1, the PDSCH itself often occupies only a fraction of this potential bandwidth \cite{golzadeh2023downlink,3gpp_38_212, 3gpp_38_213,3gpp_38_214}. Since  the PDSCH frequency allocation is mandated to  use Downlink Resource Allocation Type 1 (i.e., a contiguous set of RBs), the requirement creates an opportunity to insert auxiliary pilots into the unused portions of the $N_{\text{RB}}^\text{DL,BWP}$ RBs\cite{3gpp_38_214}. These pilots can, in turn, further enhance sensing performance and assist the DMRS within the PDSCH for more effective channel estimation.\footnote{The use of  auxiliary pilots for enhanced channel estimation is beyond the scope of this paper, which primarily focuses on their benefit to sensing performance.}

The design of the aforementioned auxiliary pilots falls within the  research domain of pilot placement in OFDM systems, which plays a pivotal role in optimizing ISAC performance. While initial research  centered on optimizing uniformly spaced patterns \cite{zhao2023reference}, the equidistant approach is known to limit the maximum unambiguous velocity and range. In order to overcome the ambiguity introduced by such sparsity, subsequent works have introduced more sophisticated designs. For instance, \cite{liu2025low} introduced a novel sensing pattern composed of multiple comb-type RS structures, where the comb sizes of the constituent structures are co-prime. Building upon a similar principle, \cite{mei2024coprime} utilized co-prime intervals between the RS to eliminate ambiguity. Furthermore, \cite{zhang2024ofdm} proposed a set of insertion criteria for OFDM RS patterns aiming at maximizing the extended ambiguity property (EAP) and applied this principle to super-resolution algorithms like 2D-MUSIC\cite{xie2021performance} and 2D-IAA\cite{stoica2009missing,roberts2010iterative}. However, these works primarily focused  on maximizing the interference-free detection region; consequently, the practical range and velocity estimation performance was not directly evaluated.

Distinct from the aforementioned pattern design methodologies, a novel approach inspired by array design theory reformulates the RS placement problem as a 2D-array design task \cite{ren2024multi}. Specifically, the proposed sparse pilot scheme achieves the same unambiguous range and peak sidelobe level (PSL) as uniform pilot placement, but with a substantial reduction in required resource elements (REs). Nevertheless, as the subsequent parameter estimation relies on a conventional 2D-FFT algorithm to generate the range-Doppler maps (RDMs), the estimation accuracy remains limited.

Motivated by the above findings, this paper proposes a novel two-stage coarse-to-fine estimation framework to leverage 5G NR broadcast signals and newly designed sparse pilots for high-precision UAV sensing. Specifically, this framework is designed to
operate within the constraints of the 5G NR protocol while
overcoming the inherent limitations of standard signals. The main contributions of this paper are summarized as follows:

\begin{itemize}
  \item First, we establish a coarse estimation stage (Stage I) that fuses information from multiple 5G NR initial access signals. By combining the  RDMs derived from the SSB and SIB1 with the high-resolution range profile from the Type\#0-PDCCH, this stage provides robust initial target detection by utilizing the area-based combination CFAR (ABC-CFAR) algorithm.
  
  \item Second, to overcome the resolution limitations of standard signals, we design a novel sparse pilot structure (SPS) inspired by sparse array theory. The SPS is strategically inserted  into the \#CORSET0 bandwidth. We then develop a corresponding high-resolution refinement algorithm (Stage II), which employs the weighted unwrapped phase (WUP) estimator and a RELAX-based iterative method to accurately extract off-grid range and velocity parameters from these sparse pilots.
  
  \item Furthermore, we introduce the density-based spatial clustering of applications with noise (DBSCAN)  based  algorithm that operates in the joint range-velocity-angle domain. This method effectively identifies and prunes redundant detections of a single target that arise from the intentional overlap of adjacent beams during the SSB/SIB1 sweep.

\item Finally, comprehensive simulation results are presented to validate the effectiveness of the proposed framework. The results demonstrate that our two-stage approach achieves superior estimation accuracy. Notably, the proposed WUP-based refinement algorithm significantly outperforms conventional benchmark methods, including 2D-FFT, 2D-MUSIC, and CS-AN, particularly under the challenging conditions imposed by the sparse pilot configuration.

\end{itemize}

\textit{Notations}: Lowercase letters, boldface lowercase letters, boldface uppercase letters, and calligraphy letters denote the scalars, vectors, matrices, and sets, respectively, i.e., $a$, $\mathbf{a}$, $\mathbf{A}$, and $\mathcal{A}$. The operators $(\cdot)^T$ and $(\cdot)^H$ represent the transpose and Hermitian transpose, respectively. The operator $\odot$ denotes the Hadamard (element-wise) product. $\mathbb{R}$ and $\mathbb{C}$ denote the sets of real and complex numbers. $[\mathbf{A}]_{i,l}$ denotes the element in the $i$-th row and $l$-th column of matrix $\mathbf{A}$. $|\cdot|$ represents the absolute value of a scalar or the cardinality of a set. $\angle(\cdot)$ denotes the angle of a complex number. $\text{Diag}(\cdot)$ returns a diagonal matrix. $\operatorname{vecl}(\cdot)$ stacks the strict lower triangular elements of a matrix into a vector. $\operatorname{round}(\cdot)$ rounds to the nearest integer. $\cup$ denotes the set union operation.

\section{SYSTEM MODEL}

\begin{figure*}[htbp]
\centerline{\includegraphics[width=0.9\textwidth]{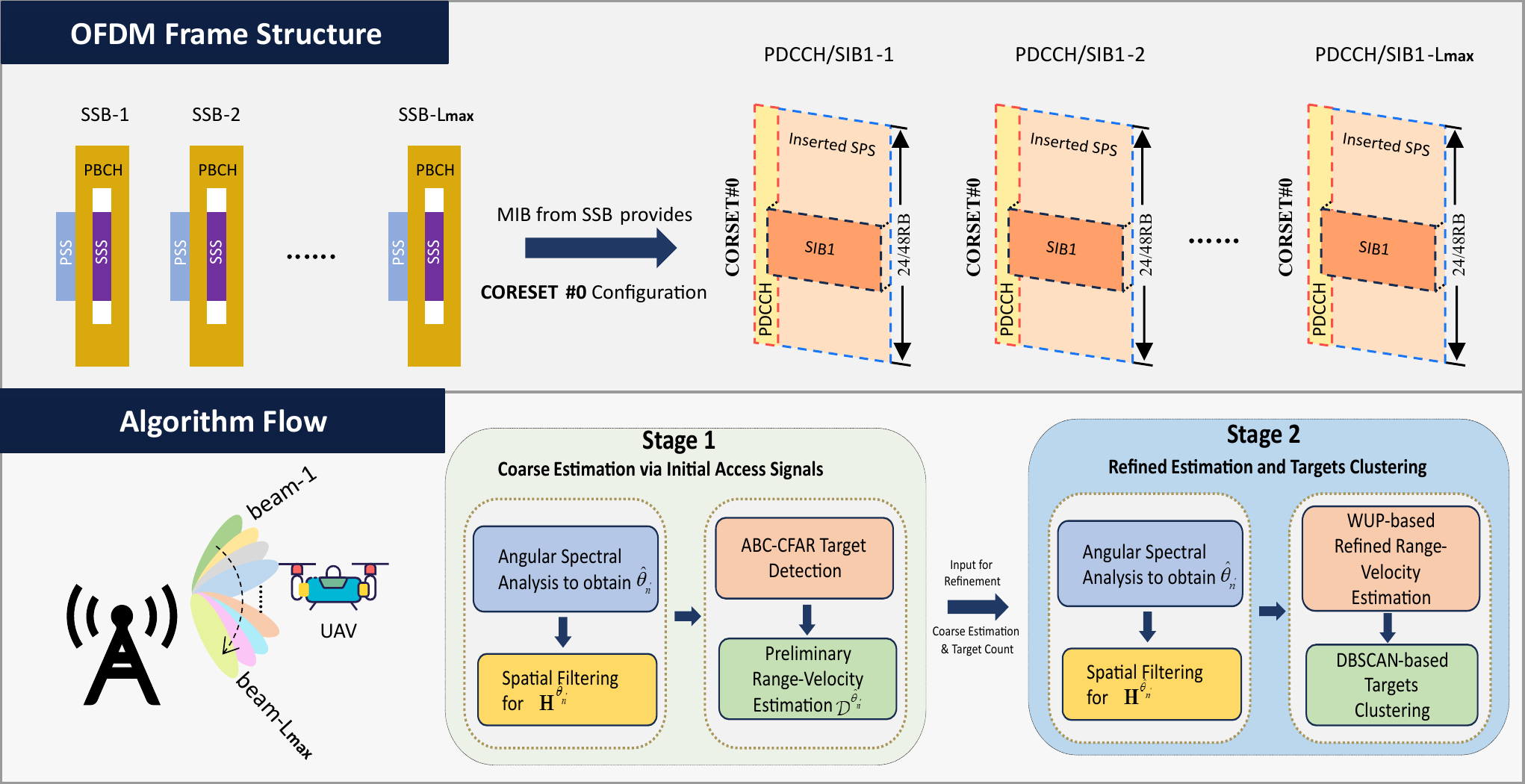}}
\caption{Frame structure and algorithm flow for the proposed two-stage  sensing framework.}
\label{fig:Flowchat}
\end{figure*}

In this paper, we consider a mono-static MIMO-OFDM ISAC system designed for simultaneous downlink communication and radar parameter estimation. The BS is equipped with separate uniform linear arrays (ULAs), consisting of $N_t$ transmit and $N_r$ receive antennas, respectively.\footnote{We assume that the BS operates under an advanced full-duplex regime, where self-interference (SI) at the receive antennas is effectively canceled.} As illustrated in Fig.~\ref{fig:Flowchat}, the  BS first transmits SSB beams to scan the coverage area, immediately followed by Type\#0-PDCCH and SIB1 signals with the same beam pattern. According to the  prior studies \cite{ksikezyk2023opportunities, golzadeh2023downlink}, these broadcast signals are utilized for a dual role. On  one hand, they deliver the information necessary for uplink RACH procedures. On the flip side, they are adopted as sensing signals. 

\subsection{Transmitted Signal Model}

Let $l\in \{1,\cdots, N_s\}$ and $i\in \{1,\cdots, N_c\}$ denote the indices for the OFDM symbols and subcarriers, respectively. Based on the frequency-flat assumption, the signal vector transmitted at the $l$-th symbol on the $i$-th subcarrier is modeled as \footnote{Following \cite{fu2023tutorial}, the signal bandwidth in this work is assumed to be contained within a single subband. Consequently, a common precoding vector is applied across all subcarriers  for a given OFDM symbol.} 
\begin{equation}
\label{eq:signal_model}
[\mathbf{X}]_{i,l} = \mathbf{w}[l] [\mathbf{S}]_{i,l},
\end{equation}
where $\mathbf{S} \in \mathbb{C}^{N_c\times N_s}$ denotes the transmitted matrix, and $\mathbf{w}[l] \in \mathbb{C}^{N_t}$ denotes the precoding vector for the $l$-th symbol. Specifically, the vector $\mathbf{w}[l]$ is selected from a predefined codebook $\mathcal{W}$ to steer the transmission towards a specific direction. This codebook, $\mathcal{W}$,  is constructed from an oversampled discrete fourier transform (DFT) grid to achieve the fine-grained angular coverage. Specifically, given an integer oversampling factor $O$, the codebook is defined as $\mathcal{W} = \{\mathbf{w}_q\}_{q=0}^{Q-1}$, where  $Q = O \cdot N_t$  and the $q$-th beamforming vector is given by
\begin{equation}
\label{eq:dft_beam_oversampled}
\mathbf{w}_{q} = \frac{1}{\sqrt{N_t}}\left[1, e^{j \frac{2 \pi q}{Q}}, e^{j \frac{2 \pi 2 q}{Q}}, \dots, e^{j \frac{2 \pi (N_t-1) q}{Q}}\right]^{T}.
\end{equation}

\noindent Then, after the inverse fast fourier transform (IFFT) operation at the transmitter,  the baseband time-domain OFDM signal can be expressed as
\begin{equation}
\label{eq:analog_signal}
\mathbf x(t) = \sum_{l=0}^{N_s-1} \sum_{i=0}^{N_{\mathrm{c}}-1} [\mathbf{X}]_{i,l} \, e^{j 2 \pi i \Delta f t} \, r(t-l T)\in \mathbb{C}^{N_t},
\end{equation}

\noindent where $\Delta f$ is the subcarrier spacing, $r (t)$
denotes the rectangular pulse, and $T=T_\text{c}+T_\text{cp}$ is the total duration of an OFDM symbol.  Here, $T_\text{c}$ denotes the OFDM symbol duration and $T_\text{cp}$ denotes the cyclic prefix (CP) duration.

\subsection{Sensing Signal Model}

 Let $\mathbf{\tilde{y}}(t) \in \mathbb{C}^{N_r \times 1}$ denote the received echoes from $N_{\text{target}}$ unknown targets. The received signal model can be written as
\begin{equation}
\label{eq:vector_signal}
\mathbf{\tilde{y}}(t) = \sum_{n=1}^{N_{\text{target}}} \beta_n e^{j 2 \pi f_{D,n} t} \mathbf{b}_\text{r}(\theta_n) \mathbf{a}_{\text{t}}^H(\theta_n) \mathbf{x}(t - \tau_n) + \mathbf{z}(t),
\end{equation}
\noindent where the components are defined as follows:
\begin{itemize}
    \item $\mathbf{\tilde{y}}(t) = [\tilde{y}(1,t), \dots, \tilde{y}(N_r,t)]^T\in \mathbb C^{N_r}$ is the received signal vector at the $N_r$ receive antennas.
    
    \item $\tau_n = 2R_n/c$ is the round-trip propagation delay for the $n$-th target.
    
    \item $\mathbf{x}(t) \in \mathbb{C}^{N_t \times 1}$ is the vector of transmitted baseband signals from the $N_t$ transmit antennas.
    
    \item $\mathbf{a}_\text{t}(\theta_n) \in \mathbb{C}^{N_t \times 1}$ is the transmit array steering vector towards the direction $\theta_n$, with $[\mathbf{a}_\text{r}(\theta_n)]_m = e^{-j 2 \pi (m-1) d_t \sin(\theta_n)/\lambda_c}$.
    
    \item $\mathbf{b}_\text{r}(\theta_n) \in \mathbb{C}^{N_r \times 1}$ is the receive array steering vector for the direction $\theta_n$, with $[\mathbf{b}_\text{r}(\theta_n)]_m = e^{-j 2 \pi (m-1) d_r \sin(\theta_n)/\lambda_c}$.
    
    \item $\mathbf{z}(t) = [z(1,t), \dots, z(N_r,t)]^T\in \mathbb C^{N_r}$  is the additive white Gaussian noise (AWGN) vector.
\end{itemize}

\noindent After the analog-to-digital conversion (ADC) with a sampling frequency of $F_\text{s}$ and the subsequent CP removal, the discrete echo signal vector for the $l$-th symbol $\mathbf{\hat{y}}(j,l) \in \mathbb{C}^{N_r \times 1}$ is obtained as
\begin{equation}
\label{eq:sampled_signal_vector}
\mathbf{\hat{y}}(j,l) \triangleq \mathbf{\tilde{y}}(lT + j/F_\text{s} + T_{\text{CP}}).
\end{equation}
Subsequently, following serial-to-parallel conversion, an $N_s$-point FFT is applied to $\mathbf{\hat{y}}(j,l)$ along the sample index dimension.  As derived in Appendix~A, the resulting frequency-domain OFDM symbol $\mathbf{y}(i,l)$ is given by
\begin{equation}
\label{eq:dft_signal_vector_b}
\mathbf{y}(i,l) = \sum_{n=1}^{N_{\text{target}}} \beta_n \mathbf{a}_\text{t}^H(\theta_n) [\mathbf{X}]_{i,l} \, e^{j i \omega_r(R_n)} e^{j l \omega_v(v_n)} \mathbf{b}_\text{r}(\theta_n) + \mathbf{z}(i,l),
\end{equation}

\noindent where $\omega_r(R_n) \triangleq -4\pi \Delta f R_n/c$ and $\omega_v(v_n) \triangleq 2\pi f_{D,n} T$ are the normalized frequencies corresponding to the range and velocity of the $n$-th target, respectively. The notation  $\mathbf{z}(i,l)$ denotes the noise term, which is approximated as AWGN\cite{pucci2022system, zhang2024target}.

\section{Stage I: Coarse Target Detection and Parameter Estimation}

 In this section, we elaborate on the utilization of the SSB signal for target detection and coarse parameter estimation. Specifically, each SSB block occupies 20 resource blocks (RBs) and four consecutive OFDM symbols in the time-frequency domain and is broadcast periodically, as illustrated in Fig.~\ref{fig:Flowchat}.  We exploit the inherent beam-sweeping nature of  SSB to obtain coarse estimates of the target's angular, range, and velocity parameters.

\subsection{Angular Spectral Analysis}
 We begin by addressing the angular estimation. To perform Direction of Arrival (DoA) estimation for targets within the $k$-th SSB beam centered at $\theta_k$, we employ the  high-resolution multiple signal classification (MUSIC) algorithm\cite{pucci2022system}. Firstly, we average the outer product of $\mathbf{y}(i,l)$ across all available subcarriers and OFDM symbols
\begin{equation}
\label{eq:scm_calculation}
\mathbf{\hat{R}} = \frac{1}{|\mathcal{R}_{k,\text{SSB}}|}  \sum_{\{(i,l)\} \in \mathcal{R}_{k,\text{SSB}} }\mathbf{y}(i,l) \mathbf{y}(i,l)^H,
\end{equation}
where  $\mathcal{R}_{k,\text{SSB}}$ is the set of REs for the $k$-th SSB beam and  $|\mathcal{R}_{k,\text{SSB}}|$ denotes the cardinality of this set.

Then, we perform the eigenvalue decomposition  as
\begin{equation}
\label{eq:eigen_decomposition}
\mathbf{\hat{R}} = \mathbf{U} \mathbf{\Lambda} \mathbf{U}^H,
\end{equation}
where $\mathbf{U} = [\mathbf{u}_1, \dots, \mathbf{u}_{N_r}]$ is the matrix of eigenvectors and $\mathbf{\Lambda}=\operatorname{Diag}(\lambda_1 \ge \lambda_2 \ge \dots \ge \lambda_{N_r})$ is the diagonal matrix containing the corresponding eigenvalues, sorted in descending order.

Based on the auto-correlation coefficient matrix algorithm\cite{salman2015estimating}, the number of potential UAVs $N_{\theta_k}$ can be effectively determined\footnote{Unlike the ideal assumption in \cite{tang2025cooperative,pucci2022system}, the limited samples per SSB beam cause information-theoretic criteria to perform poorly; MDL\cite{wax2003detection} becomes inaccurate while AIC tends to overestimate. We therefore employ the more robust auto-correlation coefficient matrix method. The interested reader is referred to \cite{salman2015estimating} for further details.}. Once $N_{\theta_k}$ is estimated, the noise subspace matrix $\bar{\mathbf{U}}$ is constructed as $\bar{\mathbf{U}} = [\mathbf{u}_{N_{\theta_k}+1}, \dots, \mathbf{u}_{N_r}].$  Then, by exploiting the orthogonality between the array steering vectors and the noise subspace, the DoA of the potential UAVs in the current beam  is given by
\begin{equation}
\label{eq:angle_estimation_argpeak}
\{\hat{\theta}_{n}\}_{n=1}^{N_{\theta_{k}}} = \underset{\theta}{\text{argpeak}} \left( \frac{1}{\mathbf{b}_\text{r}^H(\theta) \bar{\mathbf{U}} \bar{\mathbf{U}}^H \mathbf{b}_\text{r}(\theta)}, N_{\theta_k} \right),
\end{equation}
where the search window  $\theta \in [\theta_k - \Delta\theta, \theta_k + \Delta\theta]$, with  $\Delta\theta \approx 85^\circ / (N_t \cos(\theta_k))$ denoting the  approximate half-beamwidth  of the ULA at the  suppression level -10\,dB\cite{van2002optimum}\footnote{The overlap between the half-power beamwidths of adjacent SSB beams may cause a single target to be detected multiple times. However, the clustering method   is introduced in Section~V to prune  redundant detections and enhance the overall estimation accuracy.}. The operator $\text{argpeak}(\cdot,N)$ finds the locations of the $N$ largest peaks.

\subsection{Spatial Filtering}

For each estimated DoA $\hat{\theta}_{n^{\prime}}\in\{\hat{\theta}_{n}\}_{n=1}^{N_{\theta_{k}}}$ obtained from the preceding angular spectral analysis, we first perform the spatial filtering and remove the signal-dependent
coefficients $\mathbf{a}_\text{t}^H(\hat{\theta}_{n'})[\mathbf{X}]_{i,l}$\cite{xiao2024novel}. The resulting 2D channel response $[\mathbf H^{\hat{\theta}_{n^\prime}}]_{i,l}$ is given in \eqref{eq:channel_equalization_expanded_cluster}, which is shown at the top of this  page. In this model, $\mathcal{K}_{n'}$ represents the index set of targets whose DoAs are clustered around the estimated angle $\hat{\theta}_{n^{\prime}}$ . 

\begin{figure*}[t]
\normalsize
\begin{align}
\label{eq:channel_equalization_expanded_cluster}
[\mathbf H^{\hat{\theta}_{n^\prime}}]_{i,l} &= \frac{\mathbf{b}_\text{r}^H(\hat{\theta}_{n'})\mathbf{y}(i,l)}{\mathbf{a}_\text{t}^H(\hat{\theta}_{n'})[\mathbf{X}]_{i,l}} \nonumber \\
&= \frac{\mathbf{b}_\text{r}^H(\hat{\theta}_{n'}) \left( \sum_{n=1}^{N_{\text{target}}} \beta_n \mathbf{a}_\text{t}^H(\theta_n) [\mathbf{X}]_{i,l} e^{j i \omega_r(R_n)} e^{j l \omega_v(v_n)} \mathbf{b}_\text{r}(\theta_n) + \mathbf{z}(i,l) \right)}{\mathbf{a}_\text{t}^H(\hat{\theta}_{n'})[\mathbf{X}]_{i,l}}  \\
&= \underbrace{\sum_{k \in \mathcal{K}_{n'}} G_{n',k} \beta_{k} e^{j i \omega_r(R_{k})} e^{j l \omega_v(v_{k})}}_{\text{Desired Signals from Target Cluster}} \nonumber 
+ \underbrace{\sum_{n \notin \mathcal{K}_{n'}} \beta_n \frac{\mathbf{a}_\text{t}^H(\theta_n)[\mathbf{X}]_{i,l}}{\mathbf{a}_\text{t}^H(\hat{\theta}_{n'})[\mathbf{X}]_{i,l}} \left(\mathbf{b}_\text{r}^H(\hat{\theta}_{n'}) \mathbf{b}_\text{r}(\theta_n)\right) e^{j i \omega_r(R_n)} e^{j l \omega_v(v_n)}}_{\text{Interference from Other Targets}} + \underbrace{\frac{\mathbf{b}_\text{r}^H(\hat{\theta}_{n'}) \mathbf{z}(i,l)}{\mathbf{a}_\text{t}^H(\hat{\theta}_{n'})[\mathbf{X}]_{i,l}}}_{\text{Noise}}
\end{align}
\hrulefill
\end{figure*}

Due to the signal-to-noise ratio (SNR) gains introduced by the spatial filter, both the interference from other target clusters ($n \notin \mathcal{K}_{n'}$) and the noise term are significantly suppressed. Thus, the model in \eqref{eq:channel_equalization_expanded_cluster} can be  approximated by its dominant signal components
\begin{equation}
\label{eq:final_channel_model_approx}
[\mathbf H^{\hat{\theta}_{n^\prime}}]_{i,l}\approx \sum_{k \in \mathcal{K}_{n'}} \alpha_{k} e^{j i \omega_r(R_{k})} e^{j l \omega_v(v_{k})},
\end{equation}
where $\alpha_{k} = G_{n\prime,k} \beta_{k}$ is the effective complex amplitude of the $k$-th target within $\mathcal{K}_{n'}$.

\textit{\textbf{Remark1:}} In practical systems, the half-power beamwidths of  adjacent beams overlap to ensure effective user coverage. Consequently, the resource set for estimation can be expanded, e.g., $\mathcal{R}_{k,\text{SSB}}^{\text {ext }}=\mathcal{R}_{k-1,\text{SSB}} \cup \mathcal{R}_{k,\text{SSB}} \cup\mathcal{R}_{k+1,\text{SSB}}$. The expansion introduces a trade-off: a larger resource set provides more samples but suffers from lower beamforming gain at the periphery. The reduced beamfoming gain $\mathbf{a}_\text{t}^H(\hat{\theta}_{n'})[\mathbf{X}]_{i,l}$ increases the relative power of the interference and the noise terms in \eqref{eq:channel_equalization_expanded_cluster}. Therefore, the estimation window must be judiciously configured to balance the number of samples against the degradation in effective signal to interference plus noise ratio (SINR).\footnote{ Due to the beam sweeping of the SSB, a target is only coherently illuminated during a subset of OFDM symbols. Processing the entire set of symbols would incorporate noise from irrelevant time instances where the beam was pointing elsewhere, thereby degrading the effective SNR.}

\begin{figure}[ht!] 
    \centering 

    \begin{subfigure}[b]{0.15\textwidth} 
        \includegraphics[width=\linewidth]{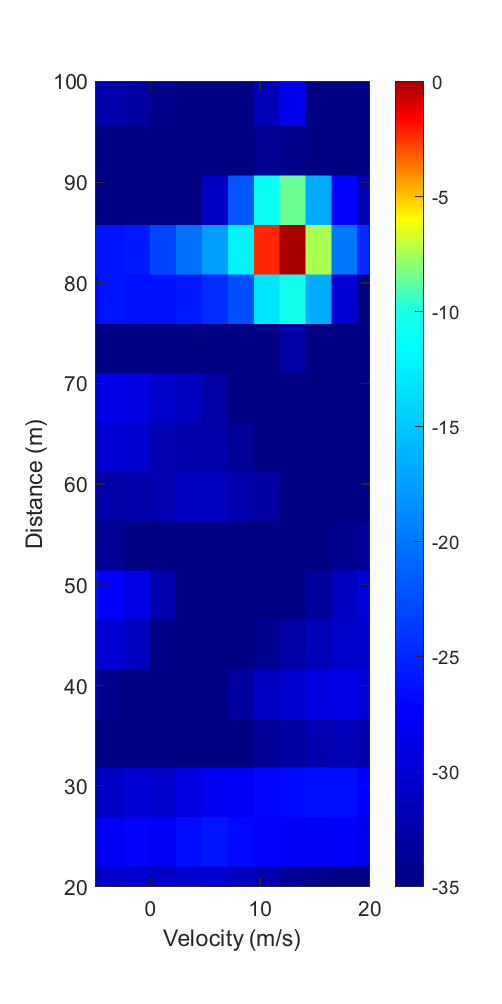}
        \caption{}
        \label{fig:sub1}
    \end{subfigure}
    \begin{subfigure}[b]{0.15\textwidth}
        \includegraphics[width=\linewidth]{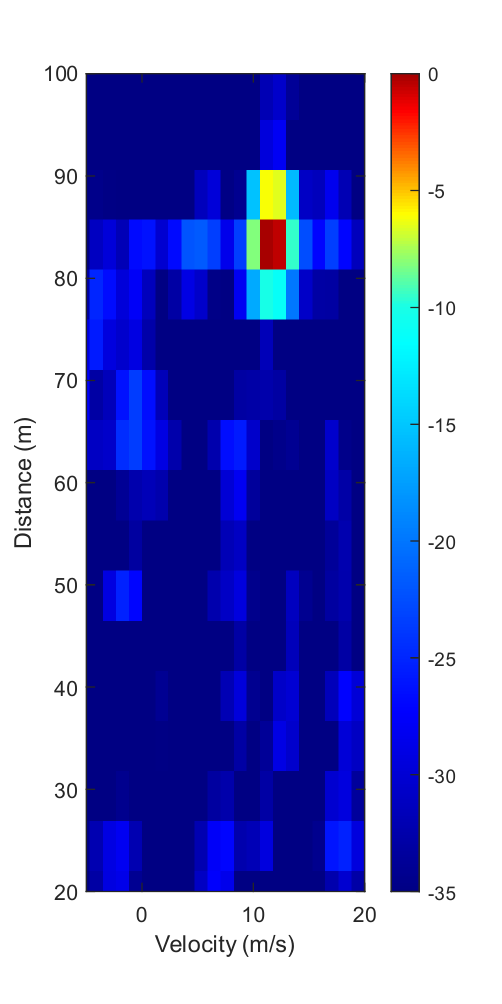}
        \caption{}
        \label{fig:sub2}
    \end{subfigure}
    \begin{subfigure}[b]{0.15\textwidth}
        \includegraphics[width=\linewidth]{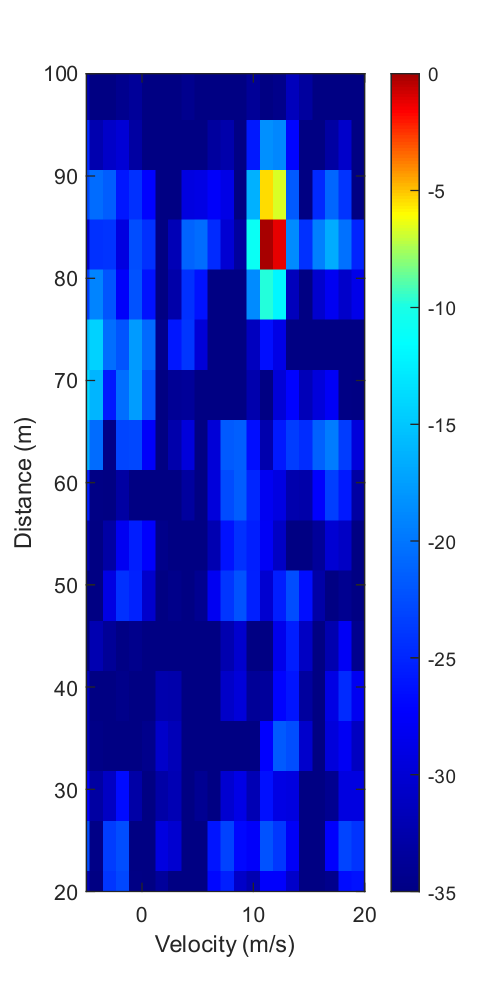}
        \caption{}
        \label{fig:sub3}
    \end{subfigure}
\caption{RDMs with different numbers of beams: (a) 1 beams, (b) 3 beams, and (c) 5 beams.}
    \label{fig:three_images}
\end{figure}

\subsection{Preliminary Range and Velocity Estimation}
Before detailing our  estimation algorithm, we first clarify two primary limitations inherent to the SSB's structure.
\begin{enumerate}[label=\arabic*)]
	\item \textbf{Limited Velocity Estimation Fidelity:} The achievable velocity resolution is fundamentally limited by the short observation interval, as a UAV target is typically illuminated by only a few four-symbol SSB blocks. 
	\item \textbf{Limited Bandwidth and Range Resolution:} The spectral bandwidth of the SSB is confined to 240 subcarriers. This restricted bandwidth inherently limits the achievable range resolution. 
\end{enumerate}

\noindent To address these limitations,  we further leverage the Type\#0-PDCCH and SIB1 broadcasting signals, which offer a more extended observation interval and wider bandwidth than SSB, to enhance velocity estimation and range resolution. Firstly, the range-Doppler map (RDM) of the SSB is  generated by performing 2D-FFT on the approximated channel matrix $\mathbf{H}^{\hat{\theta}_{n^\prime}}$ as
\begin{equation}
    [\mathbf{R}^{\hat{\theta}_{n^\prime}}_{\text{SSB}}]_{p,q} = \left| \sum_{(i,l) \in \mathcal{R}_{k,\text{SSB}}^{\text {ext }}} [\mathbf{H}]_{i,l}^{\hat{\theta}_{n^\prime}} e^{-j \frac{2\pi p i}{N_r}} e^{-j \frac{2\pi q l}{N_d}} \right|,
\end{equation}
where $N_r$ and $N_d$ are the  FFT points for the range and doppler dimensions, respectively, with zero-padding applied to improve resolution. Similarly, we  obtain the RDM of SIB1 as $\mathbf{R}^{\hat{\theta}_{n^\prime}}_{\text{SIB1}}$. The wider-bandwidth PDCCH, in turn, yields a high-resolution range profile, defined as the vector  $\mathbf{p}^{\hat{\theta}_{n^\prime}}_{\text{PDCCH}}$. These components are then fused to form the combined RDM as
\begin{equation}
    \mathbf{R}^{\hat{\theta}_{n^\prime}}_{\text{comb}} = \mathbf{R}^{\hat{\theta}_{n^\prime}}_{\text{SSB}} \odot \mathbf{R}^{\hat{\theta}_{n^\prime}}_{\text{SIB1}} \odot \mathbf{P}^{\hat{\theta}_{n^\prime}}_{\text{PDCCH}},
\end{equation}
where $\mathbf{P}^{\hat{\theta}_{n^\prime}}_{\text{PDCCH}} = \mathbf{p}^{\hat{\theta}_{n^\prime}}_{\text{PDCCH}} \mathbf{1}_{N_d}^T$, with $\mathbf{1}_{N_d}^T$ denoting a row vector of $N_d$ ones. To visualize the combined RDM, Fig.~\ref{fig:three_images} illustrates the combined RDM obtained with different numbers of beams used for angular estimation. It is clear that the 3-beam combination (Fig.~\ref{fig:three_images}b) achieves superior resolution over the single-beam case (Fig.~\ref{fig:three_images}a), while avoiding the noise-floor degradation apparent in the 5-beam scenario (Fig.~\ref{fig:three_images}c). Additionally, inaccuracies in the initial angle estimation often cause the energy of a point-based UAV target to disperse across a certain area in the RDM. Therefore, rather than the cell-average (CA) constant false-alarm rate (CFAR), we employ the area-based combination CFAR (ABC-CFAR) framework \cite{wei2022area} for target detection. The resulting set of estimated ranges and velocities for the detected targets can be expressed as
\begin{equation}
\mathcal{D}^{\hat{\theta}_{n^\prime}} = \{(\hat{R}_m, \hat{v}_m,\hat{\theta}_{n^\prime})\}_{m=1}^M,
\end{equation}
where $M$ is the total number of detected targets and $(\hat{R}_m, \hat{v}_m,\hat{\theta}_{n^\prime})$ represents the estimated range-velocity-angle pair for the $m$-th target.

\section{Sparse Pilot Structure for Refined Estimation}

In this section, we present the necessary definitions and common structures of the SPS. The design of this SPS is directly motivated by 3GPP specifications\cite{3gpp_38_211,3gpp_38_212,3gpp_38_213,3gpp_38_214} and \cite{golzadeh2023downlink}, which indicate that the PDSCH carrying SIB1 typically occupies only a portion of the CORESET\#0 bandwidth. This characteristic, combined with inspiration from 1D sparse array theory \cite{wang2024coprime}, leads to our PTRS-like SPS configuration where pilots are sparsely arranged in the frequency domain. The approach yields two key advantages: i) spanning a wide effective bandwidth to improve range resolution; ii) multiplexing communication pilots to facilitate initial access. We now introduce the necessary definitions for this SPS.

\begin{figure}[htbp]
\centerline{\includegraphics[width=0.5\textwidth]{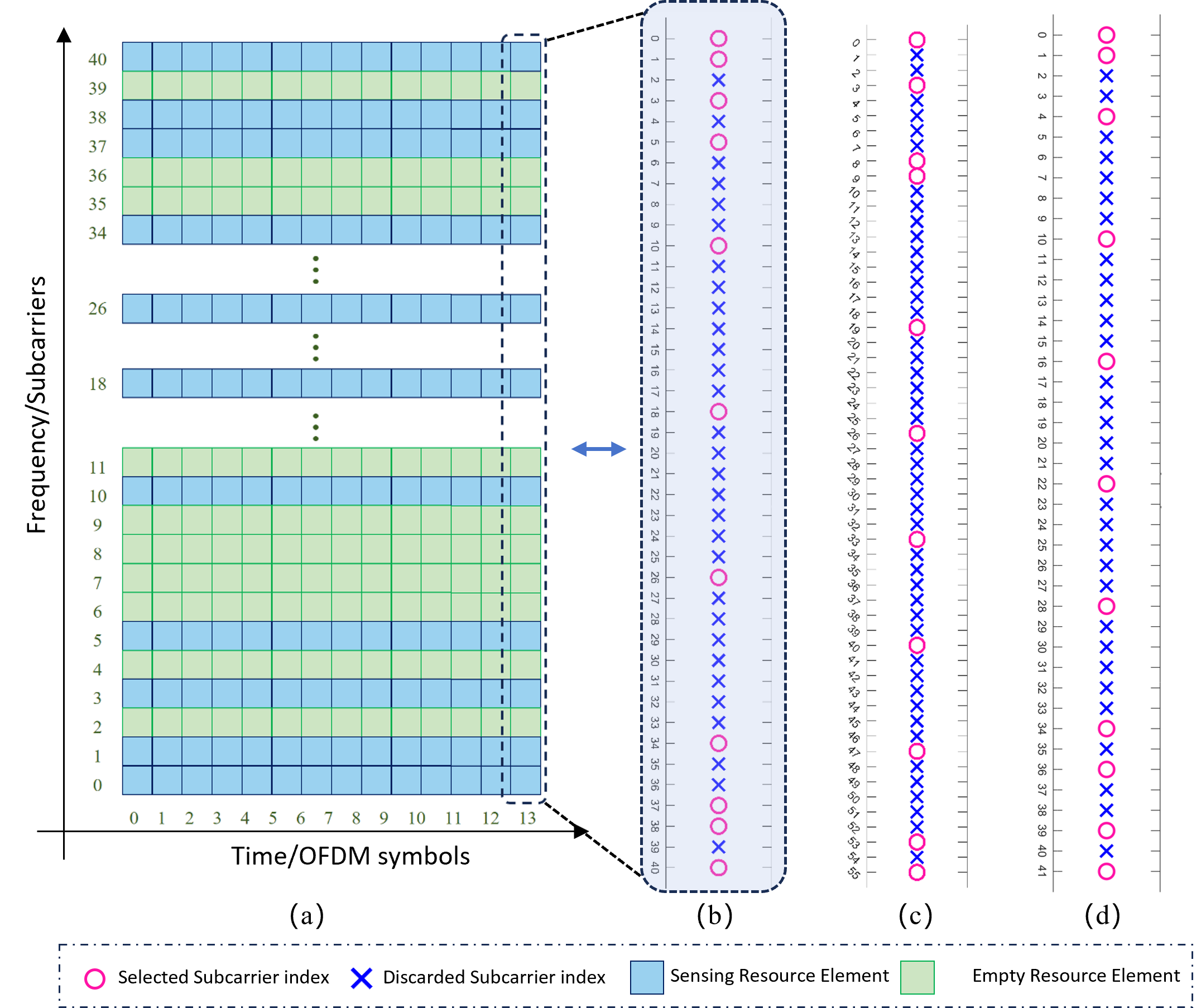}}
\caption{Example of the Sparse Pilot Structure.}
\label{fig:sensing signal structure}
\end{figure}

\begin{figure*}[htbp]
\centerline{\includegraphics[width=0.65\textwidth]{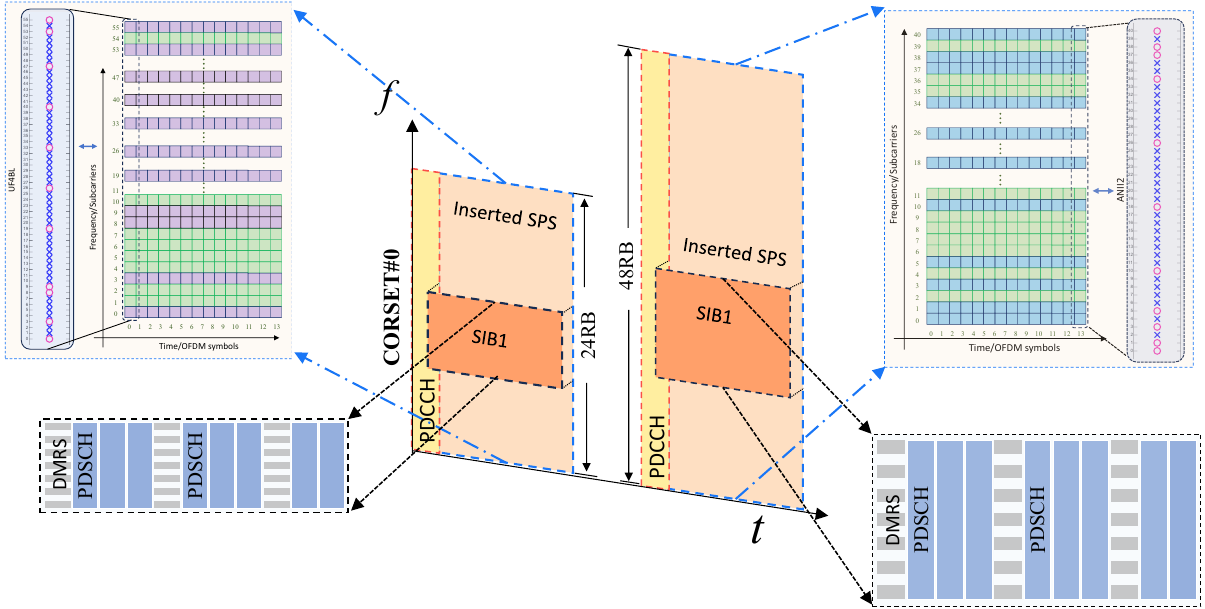}}
\caption{Example of the SPS Configuration.}
\label{fig:Inserted Sparse Signal}
\end{figure*}

\subsection{Pilots Signal Structure}

We first introduce the structure of the pilots, which are sparsely arranged in the frequency domain as illustrated in Fig.~\ref{fig:sensing signal structure}.\footnote{According to the 3GPP TS-38 series specifications, for a 120\,kHz subcarrier spacing (SCS), CORESET\#0 can be configured with a bandwidth of either 24 or 48 RBs, corresponding to 288  or 576 subcarriers, respectively. For illustrative purposes, this figure depicts a case where 11 subcarriers are occupied for sensing.}

\vspace{1ex}

\textbf{\textit{Definition 1 (Frequency-Domain Sparse Pilot Structure):}}
For a set $\mathcal{S}_f$ of subcarrier indices where pilots are placed, let $\mathcal{D}_f$ denote the corresponding Difference Co-spectrum (DCS); $\mathcal{U}_f$ denote the virtual uniform spectrum and $\mathcal{C}(k)$ denote all pilot pairs contributing to the frequency lag  $k$,
\begin{equation} \label{eq:sparse_pilot_definitions}
\begin{aligned}
    \mathcal{S}_f   &= \{p_1, p_2, \dots, p_M\}, \\
    \mathcal{D}_f &= \{p_i - p_j \mid p_i, p_j \in \mathcal{S}_f\}, \\
    \mathcal{U}_f &= [-L_u, \dots, L_u] \subseteq \mathcal{D}_f, \\
    \mathcal{C}_f(k) &= \{(p_i, p_j) \mid p_i - p_j = k; p_i, p_j \in \mathcal{S}_f\}.
\end{aligned}
\end{equation}
Here, each $p_m \in \mathcal{S}_f$ represents an individual subcarrier index, and $L_u$ is the maximum one-sided lag of the DCS $\mathcal{U}_f$. Then,  $|\mathcal{D}_f|$, $2L_u+1$, and $|\mathcal{C}_f(k)|$ are referred to the degrees of freedom (DoF), the uniform DoF (uDoF), and the weight function $w(k)$ w.r.t. the $\mathcal{S}_f$ respectively. Similar to their roles in the DoA estimation problem, it will be shown in the subsequent analysis that the DoF, uDoF, and weight function of our sparse sensing scheme are instrumental in  the performance of time-delay estimation.

\textbf{\textit{Definition 2 (ANAII-2 Sparse Pilot Structure)}}\cite{liu2017augmented}:
An Augmented Nested Array with two levels and two-sided extensions (ANAII-2) for a given total of $N$ sensors is defined by the set of sensor positions $\mathcal{S}$. First, the intermediate parameters $L$ and $M$ are determined from $N$ as
\begin{equation}
L = \text{round}(N/6), \quad M = N - 4L.
\end{equation}
The set $\mathcal{S}$ is then constructed as the union of the following five sub-sets, $\mathcal{S} = \mathcal{S}_{L11} \cup \mathcal{S}_{L12} \cup \mathcal{S}_{M} \cup \mathcal{S}_{R21} \cup \mathcal{S}_{R22}$, where each sub-set is explicitly defined by
\begin{equation} \label{eq:anaii_simplified}
\begin{aligned}
    \mathcal{S}_{M} &= \left\{ 2L^2 + L + 4Ll \mid l \in \{0, \dots, M\} \right\}, \\
    \mathcal{S}_{L12} &= \left\{ 2L^2 + L - (2L+1)l \mid l \in \{0, \dots, L-1\} \right\}, \\
    \mathcal{S}_{L11} &= \left\{ 2L+1 - l \mid l \in \{0, 2, \dots, 2L\} \cup \{2L+1\} \right\}, \\
    \mathcal{S}_{R21} &= \left\{ 2L^2 + L + 4LM + (2L-1)l \mid l \in \{0, \dots, L-1\} \right\}, \\
    \mathcal{S}_{R22} &= \left\{ 
        \begin{aligned}
            & 4L^2 - 2L + 1 + 4LM + l \mid \\
            &  l \in \{0\} \cup \{1, 3, \dots, 2L-1\} 
        \end{aligned}
    \right\}.
\end{aligned}
\end{equation}The resulting set $\mathcal{S}$ is normalized such that its minimum element is 0.

\textbf{\textit{Definition 3 (UF-4BL Sparse Pilot Structure)}}\cite{shi2021ula}:
A Uniform Fitting 4-Block-Level (UF-4BL) array with a total of $N$ sensors is defined by the set of sensor positions $\mathcal{S}$. The structure is determined by two intermediate parameters, $N_b$ and $N_t$, which are derived from $N$ as:
\begin{equation}
    N_b = \lfloor(N - 8) / 8\rfloor, \quad N_t = N - 4N_b - 6.
\end{equation}
The final set of frequency positions $\mathcal{S}$ is constructed as the union of eight distinct sub-ULAs, $\mathcal{S} = \bigcup_{i=1}^{8} \mathcal{S}_i$, where each sub-array is explicitly defined by:
\begin{align*}
    \mathcal{S}_1 &= \{ 3l \mid l = 0, 1 \} \\
    \mathcal{S}_2 &= \{ 7 + 4l \mid l = 0, 1, \dots, N_b-1 \} \\
    \mathcal{S}_3 &= \{ (4N_b + 8) + l \mid l = 0, 1 \} \\
    \mathcal{S}_4 &= \{ (4N_b + 15) + 4l \mid l = 0, 1, \dots, N_b-1 \} \\
    \mathcal{S}_5 &= \{ (8N_b + 19) + (4N_b+7)l \mid l = 0, 1, \dots, N_t-1 \} \\
    \mathcal{S}_6 &= \{ (4N_tN_b + 7N_t + 4N_b + 19) + 4l \mid l = 0, 1, \dots, N_b-1 \} \\
    \mathcal{S}_7 &= \{ (4N_tN_b + 7N_t + 8N_b + 18) + 2l \mid l = 0, 1 \} \\
    \mathcal{S}_8 &= \{ (4N_tN_b + 7N_t + 8N_b + 25) + 4l \mid l = 0, 1, \dots, N_b-1 \}
\end{align*}

\begin{table}[h!]
\centering
\caption{Comparison of Weight Functions and uDoF for Different Arrays}
\label{tab:array_comparison_final}
\begin{tabular}{l c c c c}
\toprule

\textbf{Array Type} & $w(1)$ & $w(2)$ & $w(3)$ & uDoF \\
\midrule
ANAII-2\cite{liu2017augmented} & 2 & $N/3$ & 2 & $\mathcal{O}(2N^2/3)$ \\
MISC\cite{zheng2019misc}    & 1 & $2\lfloor N/4 \rfloor - 2$ & $1/2$ & $\mathcal{O}( N^2/2 + 3N)$ \\
TSENA\cite{ren2020extended}  & $N_1 - 6$ & $N_1 - 8$ & $N_1 - 9$ & $\mathcal{O}( N^2/2 + 3.5N)$ \\
UF-3BL\cite{shi2021ula}  & 1 & 1 & $3N_b - 1$ & $\mathcal{O}( N^2/2 + 2N)$ \\
UF-4BL\cite{shi2021ula}   & 1 & 1     & 2 & $\mathcal{O}( N^2/2 + 2N)$ \\
\bottomrule
\end{tabular}

\parbox{\columnwidth}{\footnotesize \vspace{1ex} 
\textit{Note:} For  TS-ENA array, the parameter  $N_1 = 2\lfloor (N+3)/4 \rfloor$. For UF-3BL, the parameter  $N_b = \lfloor(N - 5) / 6\rfloor$.}
\end{table}

Based on the comparison in Table~\ref{tab:array_comparison_final}, we recommend the UF-4BL structure for the 24RB configuration and ANAII-2 for the 48RB configuration, respectively. For the 24RB  case, both structures offer comparable uDoF, making the UF-4BL preferable due to its simpler weight function and lower coupling. Conversely, for the 48RB case, the uDoF of ANAII-2 significantly surpasses that of UF-4BL. This substantial uDoF gain outweighs its higher coupling, rendering it the superior choice for performance.

\subsection{Integration of the Pilots and SIB1 }


In this subsection, we describe how the aforementioned SPS is incorporated into the SIB1 block. Let us denote the SIB1 data symbol matrix and the sparse sensing pilot matrix as $\mathbf{S}_{\text{SIB1}}\in \mathbb{C}^{N_{c,\text{SIB1}} \times N_{s,\text{SIB1}}}$ and $\mathbf{S}_{\text{sen}} \in \mathbb{C}^{N_{c,\text{SIB1}} \times N_{s,\text{SIB1}}}$, respectively. The placement of these signals is governed by their corresponding binary masks $\boldsymbol{\mathcal{L}}_{k,\text{SIB1}}$ and $\boldsymbol{\mathcal{L}}_{k,\text{sen}}$, which belong to the set $\{0,1\}^{N_{c,\text{SIB1}} \times N_{s,\text{SIB1}}}$. For the sensing pilot mask $\boldsymbol{\mathcal{L}}_{k,\text{sen}}$, we adopt a configuration analogous to the PT-RS in 5G NR. Specifically, the pilots occupy all OFDM symbols within the SIB1 block but are placed only on the sparse set of subcarrier indices $\mathcal{S}_f$, as described in Section~IV-A.

The final transmitted  matrix $\mathbf{S}_{\text{tx}} \in \mathbb{C}^{N_{c,\text{SIB1}} \times N_{s,\text{SIB1}}}$ is formed by having the SIB1 symbols overwrite the sensing pilots on any shared REs. This is expressed as
\begin{equation} \label{eq:signal_composition}
    \mathbf{S}_{\text{tx}} = \mathbf{S}_{\text{SIB1}} + \mathbf{S}_{\text{sen}} \odot (\mathbf{J} - \boldsymbol{\mathcal{L}}_{k,\text{SIB1}}),
\end{equation}
where $\mathbf{J}\in\mathbb{C}^{N_{c,\text{SIB1}} \times N_{s,\text{SIB1}}}$ is an all-ones matrix of the same size. The overwrite mechanism is illustrated in Fig.~\ref{fig:Inserted Sparse Signal}, where SIB1 symbols take precedence over sensing pilots. Here, we denote the set of carrier frequencies jointly occupied by both SIB1 and the sensing pilots as $\mathcal S_{f,\text{combined}}$.

\section{Stage II: Refined Parameter Estimation with Sparse Pilot Symbols}

In this section, we refine the coarse parameter estimates obtained in Stage I by applying the RELAX-based WUP algorithm. Then, the parameter pairing is introduced to identify the most probable range-velocity pairings. Finally, we employ the DBSCAN algorithm to prune redundant targets and obtain the final estimates.

\subsection{Weighted Unwrapped Phase Velocity Estimation}

We begin by considering the channel response matrix $\mathbf{H}^{\hat{\theta}_{n^\prime}}$ associated with a specific angle $\hat{\theta}_{n^\prime}$, assuming it corresponds to a single target $k$. For $i \in \mathcal{S}_{f,\text{combined}}$, we define the Doppler vector\footnote{For notational simplicity, we omit the subcarrier index $i$ dependence in the following derivations, as the process is identical for each $i \in \mathcal S_{f,\text{combined}}$.} 
\begin{equation}
    \mathbf{a}_D(v_k) = [\mathbf{H}^{\hat{\theta}_{n^\prime}}]_{i,:} = \bar{\alpha}_k [e^{j 0 \omega_v(v_k)}, \dots, e^{j (N_{s,\text{SIB1}}-1) \omega_v(v_k)}].
\end{equation}
Then, the Doppler covariance vector $\mathbf{r}^{f_D} \in \mathbb{C}^{N_{\text{vec}} \times 1}$ is constructed as
\begin{equation}
    \mathbf{r}^{f_D} = \text{vecl}(\mathbf{a}_D(v_k)^H \mathbf{a}_D(v_k)),\label{eq:spatial_covariance_vector_def}
\end{equation}
where $N_{\text{vec}} = N_{s,\text{SIB1}}(N_{s,\text{SIB1}}-1)/2$. Its $n$-th element is given by $[\mathbf{r}^{f_D}]_n = |\bar{\alpha}_k|^2 e^{j(n-C_n)\omega_v(v_k)}$, where $C_n = \max\{C_t | t \geq 0 \text{ and } C_t < n\}$ and $C_t = t N_{s,\text{SIB1}} - t(t+1)/2$.

Denoting the wrapped phase vector as $\mathbf{p} = \angle \mathbf{r}^{f_D}\in \mathbb{C}^{1 \times N_{s,\text{SIB1}}}$, its $n$-th element is $[\mathbf{p}]_n = (n-C_n) \omega_v(v_k) \pmod{2\pi}$. Based on the coarse velocity estimate $\omega_v(v_{\text{FFT}})$ obtained in Section III, we define the reference unwrapped phase vector $\mathbf{p}_{\text{FFT}}\in \mathbb{C}^{N_{\text{vec}} \times 1}$ with its $n$-th element given by
\begin{equation}
    [\mathbf{p}_{\text{FFT}}]_n = (n-C_n)\omega_v(v_{\text{FFT}}).
\end{equation}
As such, the integer wrap vector $\boldsymbol{\kappa}$, whose elements represent the number of phase wraps, can be estimated by
\begin{equation}
    \boldsymbol{\kappa} = \text{round}\left( \frac{\mathbf{p}_{\text{FFT}} - \mathbf{p}}{2\pi} \right),
\end{equation}
where $\text{round}(\cdot)$ is an element-wise rounding operation. The true unwrapped phase vector $\tilde{\mathbf{p}}$ is then recovered as
\begin{equation}
    \tilde{\mathbf{p}} = \mathbf{p} + 2\pi \boldsymbol{\kappa}.
\end{equation}

\noindent Finally, given the unwrapped phase vector $\tilde{\mathbf{p}}$, the estimate of the Doppler frequency $\hat{\omega}_v(v_k)$ is obtained as
\begin{equation}
    \hat{\omega}_v(v_k) = \frac{\sum_{n=1}^{N_{\text{vec}}} (n-C_n) [\tilde{\mathbf{p}}]_n}{\sum_{n=1}^{N_{\text{vec}}} (n-C_n)^2}.
 \label{equ:estimated_velocity}
\end{equation}

\textit{\textbf{Remark 2:}} Observing \eqref{eq:spatial_covariance_vector_def}, we note an analogy between the phase differences captured by the cross-correlation operation and the virtual difference co-arrays exploited in DoA estimation\cite{ma2021multi, liu2017augmented}. As depicted in Fig.~\ref{fig:sensing signal structure}, the considered PT-RS style reference signal structure, which is dense across OFDM symbols, is analogous to the ULA. Furthermore, the weighted unwrapped estimator in \eqref{equ:estimated_velocity} inherently assigns higher weights to phase differences corresponding to larger effective time separations (i.e., larger values of $n-C_n$), as these generally provide finer velocity resolution. This weighting strategy aligns conceptually with the objectives in the non-ULA  design, which often aims for higher DoF and lower mutual coupling.

\subsection{RELAX-Based Multi-Target Velocity Estimation}

We now extend the estimation framework to the multi-target scenario, where multiple targets may share the same angle bin $\hat{\theta}_{n^\prime}$. In this case, the $i$-th row of the channel response matrix is the superposition of $|\mathcal{K}_{n'}|$  signals :
\begin{equation}
    [\mathbf{H}^{\hat{\theta}_{n^\prime}}]_{i,:} = \sum_{k=1}^{|\mathcal{K}_{n'}|} \alpha_k [e^{j 0 \omega_v(v_k)}, \dots, e^{j (N_{s,\text{SIB1}}-1) \omega_v(v_k)}].
\end{equation}
To estimate the parameters of each target individually, we adopt the estimate-subtract strategy inspired by the RELAX algorithm \cite{ma2021multi}. This iterative method estimates the parameters of one target at a time while treating the others as interference. Specifically, assuming that the parameters of $Q-1$ targets have already been estimated in $\mathbf{H}^{\hat{\theta}_{n^\prime}}$ , the signal component corresponding to the $k$-th target ($k \le Q$) is extracted as
\begin{equation}
    \mathbf{s}_{f_D}^{(k)} = [\mathbf{H}^{\hat{\theta}_{n^\prime}}]_{i,:} - \sum_{\bar{k}=1, \bar{k} \neq k}^{Q} \hat{\alpha}_{\bar{k}} [e^{j 0 \hat{\omega}_v(v_{\bar{k}})}, \dots, e^{j (N_{s,\text{SIB1}}-1) \hat{\omega}_v(v_{\bar{k}})}],
\end{equation}
where $\hat{\alpha}_{\bar{k}}$ and $\hat{\omega}_v(v_{\bar{k}})$ are the current estimates for the other targets. The Doppler frequency $\hat{\omega}_v(v_k)$ is then estimated from $\mathbf{s}_{f_D}^{(k)}$ using the weighted unwrapped phase (WUP) method described previously. Subsequently, the complex amplitude $\alpha_k$ can be estimated as
\begin{equation}
    \hat{\alpha}_k = \frac{\mathbf{a}_D(\hat{\omega}_v(v_k))^* \mathbf{s}_{f_D}^{(k)}}{N_{s,\text{SIB1}}},
\end{equation}where $\mathbf{a}_D(\hat{\omega}_v(v_k))$ is the estimated Doppler vector for the $k$-th target's Doppler frequency. This process is iterated, refining the estimates for all targets until  all $Q = |\mathcal{K}_{n'}|$ targets have been identified. The detailed implementation is presented in Algorithm~1.


\begin{algorithm}
\caption{RELAX-Based Iterative Velocity Estimation}
\label{alg:relax_velocity_correct_final_v2}
\begin{algorithmic}[1] 
    \STATE \textbf{Input:} $\mathbf{s}_{f_D} = [\mathbf{H}^{\hat{\theta}_{n^\prime}}]_{i,:}$, $|\mathcal{K}_{n'}|$, $\tau$.
    \STATE \textbf{Initialization:} Obtain initial estimates $\{\hat{\omega}_v(v_k)\}_{k=1}^{|\mathcal{K}_{n'}|}, \{\hat{\alpha}_k\}_{k=1}^{|\mathcal{K}_{n'}|}$, $\Delta = \infty$, $\mathbf{y}_1 = \mathbf{s}_{f_D}$.

    \FOR{$Q = 1$ to $|\mathcal{K}_{n'}|$}
        \STATE Estimate $\hat{\omega}_v(v_Q)$ from $\mathbf{y}_Q$ according to (20)-(24).
        \STATE Estimate $\hat{\alpha}_Q$ from  $\mathbf{y}_Q$ and $\hat{\omega}_v(v_Q)$ according to (28).
        \STATE Reset convergence check $\Delta = \infty$.
        \WHILE{$\Delta \ge \tau$}
            \STATE Store previous estimates $\{\omega_{\text{prev}}(v_k) = \hat{\omega}_v(v_k)\}_{k=1}^{Q}$
            \FOR{$k = 1$ to $Q$}
                \STATE Calculate residual signal $\mathbf{y}_k^{\text{refine}} = \mathbf{s}_{f_D} - \sum_{\substack{\bar{k}=1 , \bar{k} \neq k}}^{Q} \hat{\alpha}_{\bar{k}} \mathbf{a}_D(\hat{\omega}_v(v_{\bar{k}}))$
                \STATE Refine velocity $\hat{\omega}_v(v_k)$ from $\mathbf{y}_k^{\text{refine}}$ according to (20)-(24).
                \STATE Refine amplitude $\hat{\alpha}_k$ from $\mathbf{y}_k^{\text{refine}}$ and $\hat{\omega}_v(v_k)$ according to (28).
            \ENDFOR
            \STATE Calculate maximum change $\Delta = \max_{k \in \{1..Q\}} \{ ||\hat{\omega}_v(v_k) - \omega_{\text{prev}}(v_k)|| \}$
        \ENDWHILE
        \STATE Prepare residual for the next target's initial estimation
        \STATE $\mathbf{y}_{Q+1} = \mathbf{s}_{f_D} - \sum_{\bar{k}=1}^{Q} \hat{\alpha}_{\bar{k}} \mathbf{a}_D(\hat{\omega}_v(v_{\bar{k}}))$
    \ENDFOR

    \STATE \textbf{Output:} Estimated velocities $\{\hat{\omega}_v(v_k)\}_{k=1}^{|\mathcal{K}_{n'}|}$.
\end{algorithmic}
\end{algorithm}

\subsection{Range Estimation and Parameter Pairing}

Drawing inspiration from research in the non-ULA design, we adopt the SPS for the pilots, as detailed in Section~IV. Let $\mathcal{S}_{f, \text{combined}} = \{p_1, \dots, p_{|\mathcal{S}_{f, \text{combined}}|}\}$ denote this sparse set of subcarrier indices. Assuming a single target $k$, we define the delay vector $\mathbf{a}_\tau(R_k) \in \mathbb{C}^{|\mathcal{S}_{f, \text{combined}}| \times 1}$ by extracting the channel response at the sparse subcarrier indices over a specific OFDM symbol $l$
\begin{equation}
    \mathbf{a}_\tau(R_k) \approx \bar{\alpha}_k [e^{j p_1 \omega_r(R_k)}, \dots, e^{j p_{|\mathcal{S}_{f, \text{combined}}|} \omega_r(R_k)}]^T.
\end{equation}
Following the weighted unwrapped phase estimator framework proposed in Sections V.A and V.B, the delay  covariance vector $\mathbf{r}^{\tau}$ is constructed as
\begin{equation}
    \mathbf{r}^{\tau} = \text{vecl}(\mathbf{a}_\tau(R_k) \mathbf{a}_\tau(R_k)^H).
\end{equation}
Then, the corresponding  unwrapped phase vector $\tilde{\mathbf{p}}^{\tau}$ can be expressed as
\begin{equation}
    \tilde{\mathbf{p}}^{\tau} = \text{vecl}(\mathbf{R}_{\tau}) \omega_r(R_k),
\end{equation}
where $[\mathbf{R}_{\tau}]_{i,j} = p_i - p_j$, with $p_i$ and $p_j$ being the $i$-th and $j$-th frequency indices from $\mathcal{S}_{f, \text{combined}}$, respectively. Notably, the vector $\text{vecl}(\mathbf{R}_{\tau})$ contains the frequency index differences corresponding to the DCS, which is detailed in Definition 1. The subsequent steps for  $\omega_r(R_k)$ proceed analogously to the velocity estimation.

Building upon the range and velocity estimation procedures outlined previously, we can obtain the sets of estimated normalized range frequencies $\{\hat{\omega}_r(R_k)\}_{k=1}^{|\mathcal{K}_{n'}|}$ and Doppler frequencies $\{\hat{\omega}_v(v_k)\}_{k=1}^{|\mathcal{K}_{n'}|}$. Next, parameter pairing is performed to associate the correct range and velocity estimates for each target. To this end, we first construct  $\mathbf{\Gamma}$ containing all possible $|\mathcal{K}_{n'}|^2$ range-velocity pairings
\begin{equation}
    \mathbf \Gamma = \{ \left(\hat{\omega}_r(R_i), \hat{\omega}_v(v_j)\right) \}_{i,j=1}^{|\mathcal{K}_{n^\prime}|},
\end{equation}
where for each pair $(\hat{\omega}_r(R_i), \hat{\omega}_v(v_j)) \in \mathbf  \Gamma$, the matching score is calculated as
\begin{equation}
    S^{\hat{\theta}_{n^\prime}}_{\hat{\omega}_r(R_i), \hat{\omega}_v(v_j)} = \left| \mathbf{a}_\tau(\hat{R}_i)^H \mathbf{H}^{\hat{\theta}_{n^\prime}} \mathbf{a}_D(\hat{v}_j) \right|.
\end{equation}
Subsequently, by identifying the $|\mathcal{K}_{n'}|$ pairs in $\mathbf \Gamma$ that yield the highest matching scores, the final set $ \mathbf \Xi^{{\hat{\theta}_{n^\prime}}}$ of paired range and velocity parameters for the $|\mathcal{K}_{n'}|$ targets is given by
\begin{equation}
    \mathbf \Xi^{{\hat{\theta}_{n^\prime}}} = \{ (\hat{R}_k, \hat{v}_k) \}_{k=1}^{|\mathcal{K}_{n'}|}.
\end{equation}

\subsection{DBSCAN-based Repeated Targets Pruning}

In order to maintain continuous energy coverage, the half-power beamwidths of adjacent beams typically overlap during SSB and SIB1 beam sweeping. Consequently, a single UAV target may be detected by multiple beams. To address this redundancy, we employ the DBSCAN algorithm \cite{ester1996density}  to cluster the detected targets in the range-velocity-angle space. To mitigate the impact of spurious noise detections, the minimum number of points $N_{\text{min}}$ in a cluster is set to 2.

As a prerequisite for clustering, we define the detected targets set $\mathbf{\Xi}$ across all detected angles $\{\hat{\theta}_{n^\prime}\}$ as
\begin{equation}
    \mathbf{\Xi} = \bigcup_{\hat{\theta}_{n^\prime}} \{ (\hat{R}_k, \hat{v}_k, \hat{\theta}_{n^\prime}) \mid (\hat{R}_k, \hat{v}_k) \in \mathbf{\Xi}^{{\hat{\theta}_{n^\prime}}} \},
\end{equation}where each element $\mathbf{s} \in \mathbf{\Xi}$ is a tuple $\mathbf{s} = (\hat{R}, \hat{v}, \hat{\theta})$. Next, we define the weighting matrix $\mathbf{W} = \text{Diag}(w_R, w_v, w_\theta)$, where the weights reflect the inverse squared resolutions: $w_R = (2 N_{c,\text{SIB1}} \Delta f / c)^2$, $w_v = (2 N_{s,\text{SIB1}} T_{\text{sym}} / \lambda)^2$ and $w_\theta \propto N_R^2$. Then, the  weighted distance $d(\mathbf{s}_1, \mathbf{s}_2)$ between two points $\mathbf{s}_1, \mathbf{s}_2 \in \mathbf{\Xi}$ is calculated as
\begin{equation}
    d(\mathbf{s}_1, \mathbf{s}_2) = \left((\mathbf{s}_1 - \mathbf{s}_2)^T \mathbf{W} (\mathbf{s}_1 - \mathbf{s}_2)\right)^{1/2},
\end{equation}
where $(\mathbf{s}_1 - \mathbf{s}_2)$ represents the element-wise difference vector.

After applying DBSCAN with the distance metric and a suitable neighborhood radius $\epsilon$, the algorithm groups the points in $\mathbf{\Xi}$ into clusters. Each resulting cluster ideally corresponds to multiple detections of a single, unique UAV target, while points classified as noise by DBSCAN are discarded. Specifically for each cluster $\mathcal{C}_m$ identified by DBSCAN, the final  estimate $(\hat{R}_m^{\text{final}}, \hat{v}_m^{\text{final}}, \hat{\theta}_m^{\text{final}})$ for the $m$-th target is obtained as
\begin{equation}
    (\hat{R}_m^{\text{final}}, \hat{v}_m^{\text{final}}, \hat{\theta}_m^{\text{final}}) = \frac{\sum_{\mathbf{s} \in \mathcal{C}_m} S_{\hat{\omega}_r(R), \hat{\omega}_v(v)}^{\hat{\theta}} \cdot \mathbf{s}}{\sum_{\mathbf{s} \in \mathcal{C}_m} S_{\hat{\omega}_r(R), \hat{\omega}_v(v)}^{\hat{\theta}}},
\end{equation}
where the summation and division are performed element-wise for the tuple $\mathbf{s} = (\hat{R}, \hat{v}, \hat{\theta})$. This process yields the final pruned  set of detected UAV parameters.

\section{Numerical Results}

In this section, simulation results are presented to evaluate the performance of the proposed sensing framework. Firstly, we provide the simulation parameter settings. Then, we compare the performance of our proposed RELAX-based WUP algorithm with conventional techniques in a single-target scenario. Furthermore, we also evaluate the impact of  SPS configurations. Finally, we compare  our DBSCAN-based clustering algorithm with the benchmark pruning method in the multi-target scenario.

\subsection{Simulation Settings}

Unless stated otherwise, the simulation parameters are set as follows: We consider a DFRC BS operating at a carrier frequency of $f_c=30$ GHz with  $\Delta f=120$ kHz. The BS is equipped with seperate half-wavelength ULAs with $N_t=32$ and $N_r=64$ antennas, respectively. Assuming the normal CP, the 0th and 7th OFDM symbols have a CP duration of $T_\text{cp} \approx 1.107  ~\mu\text{s}$ with $T \approx 9.44 ~\mu\text{s}$, while the remaining 12 symbols have $T_\text{cp} \approx 0.586 ~\mu\text{s}$ with $T \approx 8.92 ~\mu\text{s}$. The RCS of each UAV is set to $\sigma = 0.05 \mathrm{~m}^2$.  Moreover, the sensing procedure is sequential, comprising two distinct stages. Firstly, $64$ SSB beams are transmitted to perform initial detection across the $[-45^\circ, 45^\circ]$ angular range. Secondly, SPS-augmented PDCCH/SIB1 signals are transmitted to provide the essential information for UE system access, while simultaneously enabling refined parameter estimation. All results are averaged over 200 Monte Carlo trials.

\subsection{Estimation Accuracy and Computational Cost for Single Target}

We first evaluate the performance of our proposed algorithm in the single-target scenario by comparing it against  three benchmark methods\footnote{To ensure reproducibility, the source code for all benchmark algorithms is provided at \url{https://github.com/OnePieceofCakeforYou}.}:  2D-FFT\cite{pucci2022system}, 2D-MUSIC\cite{xie2021performance},  2D CS-AN\cite{zheng2017super}. These benchmarks  represent the periodogram, subspace-based and compressed sensing methods, respectively. For this simulation, the range, radial velocity, and angle of the single target are set to (50~m, 25~m/s, 15$^{\circ}$).

\begin{figure}[h]
\centerline{\includegraphics[width=0.4\textwidth]{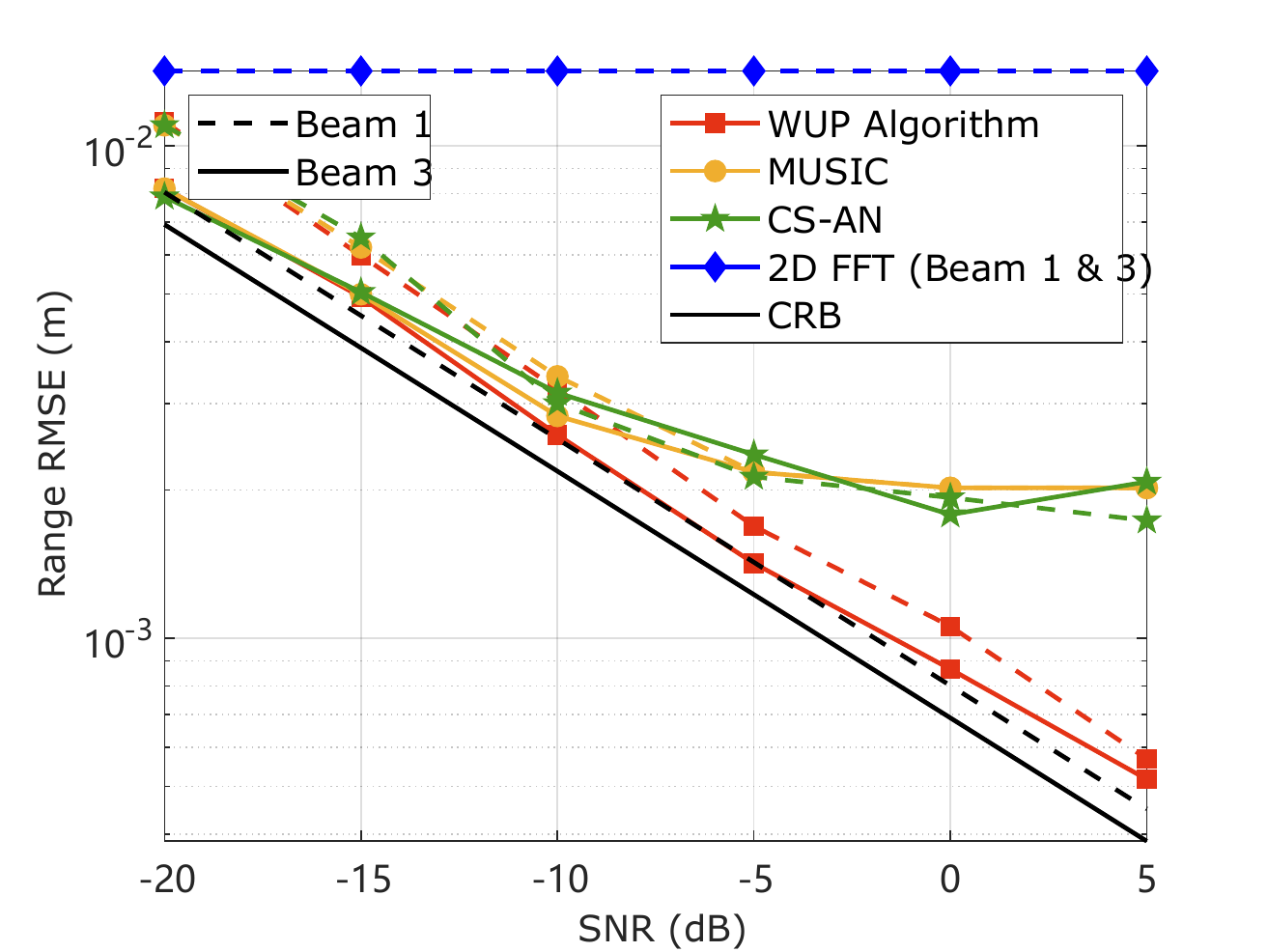}}
\caption{RMSE of range estimation versus SNR.}
\label{fig:Single_RMSE_R}
\end{figure}

Fig. \ref{fig:Single_RMSE_R} compares the range RMSE performance versus SNR for different algorithms. We can observe that the 2D-FFT method exhibits a almostly fixed RMSE, limited by bandwidth resolution and unaffected by the number of beams. In contrast, both 2D-MUSIC and CS-AN show performance gains as SNR increases; however, their grid-based search leads to eventual error floors. Among these, CS-AN provides better range estimation than 2D-MUSIC. The proposed WUP algorithm significantly outperforms all benchmarks, achieving the lowest RMSE and highlighting its effective off-grid estimation capability under challenging sparse pilot conditions.

\begin{figure}[h]
\centerline{\includegraphics[width=0.4\textwidth]{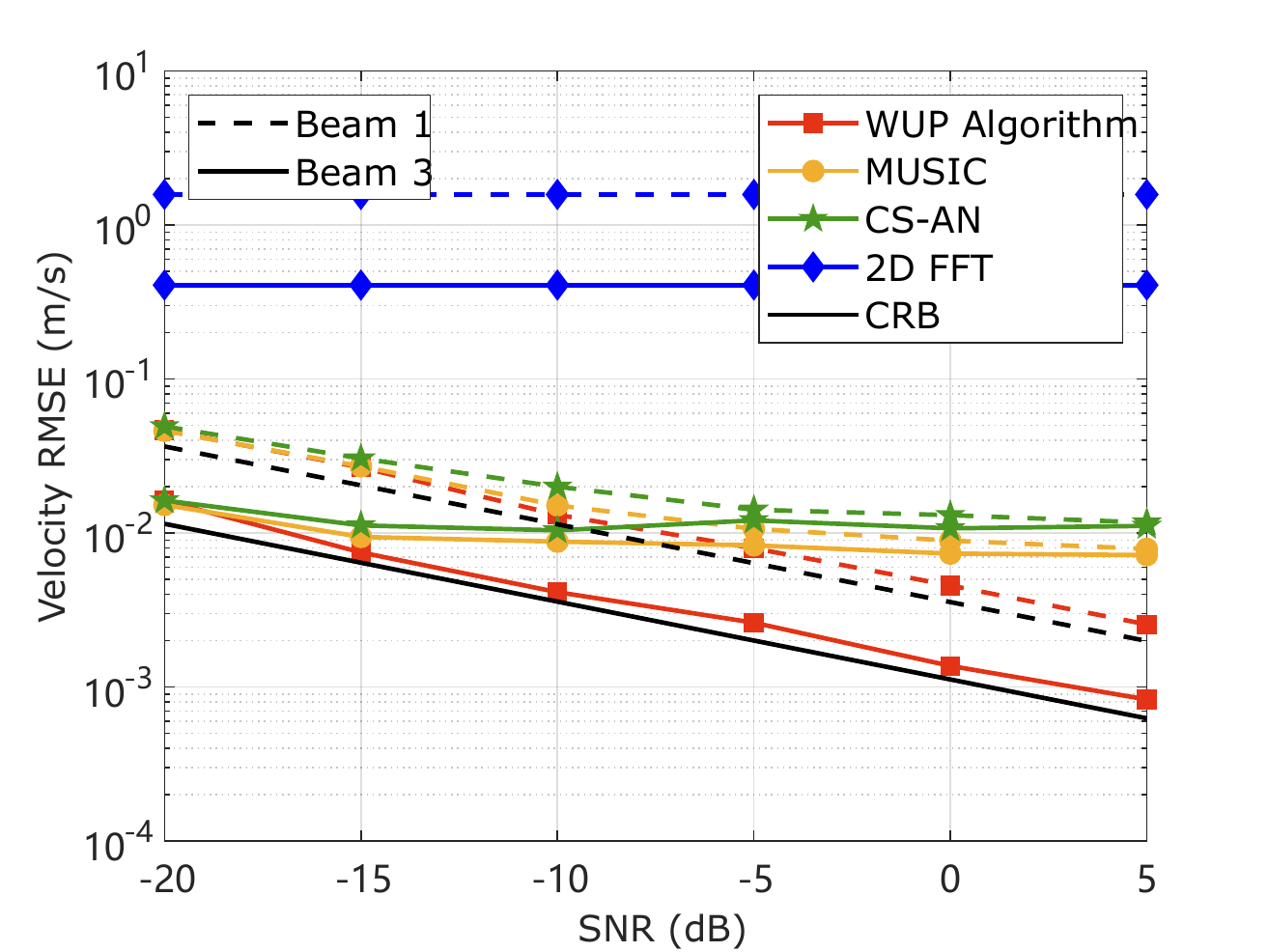}}
\caption{RMSE of velocity estimation versus SNR.}
\label{fig:Single_RMSE_V}
\end{figure}

Fig. \ref{fig:Single_RMSE_V} presents the velocity RMSE versus SNR for different algorithms. A key observation is that utilizing  multiple beams significantly enhances velocity estimation accuracy for all algorithms by extending the effective observation interval and thus improving velocity resolution. Notably, this performance gain is markedly stronger than that observed for range estimation. In contrast to the range estimation results, it is noted that the 2D-MUSIC slightly outperforms CS-AN in velocity estimation under the sparse frequency configuration. Furthermore, the proposed WUP algorithm demonstrates superior performance, consistently achieving the lowest RMSE across the entire simulated SNR range.

\begin{figure}[h]
\centerline{\includegraphics[width=0.4\textwidth]{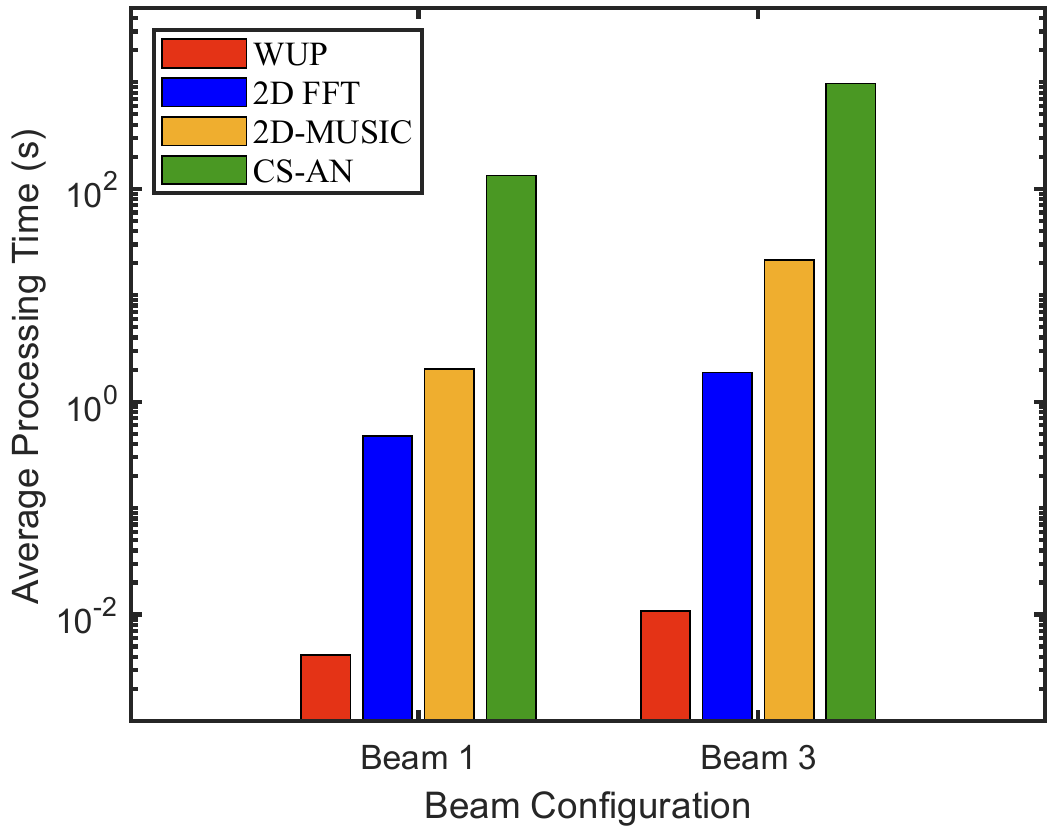}}
\caption{Computational time comparison .}
\label{fig:Single_RMSE_Avg_time}
\end{figure}

Fig. \ref{fig:Single_RMSE_Avg_time} demonstrates the average processing time  of the proposed WUP algorithm against benchmark methods under different beam configurations. The results reveal a clear computational hierarchy, with WUP algorithm being the most efficient, followed in order by 2D-FFT, 2D-MUSIC, and the substantially more costly CS-AN. While the computational burden increases for all algorithms with more beams, the WUP algorithm maintains a marked efficiency edge. This advantage demonstrates its superior scalability compared to the other methods, particularly the highly complex CS-AN.

\begin{figure}[ht!] 
    \centering 

    \begin{subfigure}[b]{0.23\textwidth} 
        \includegraphics[width=\linewidth]{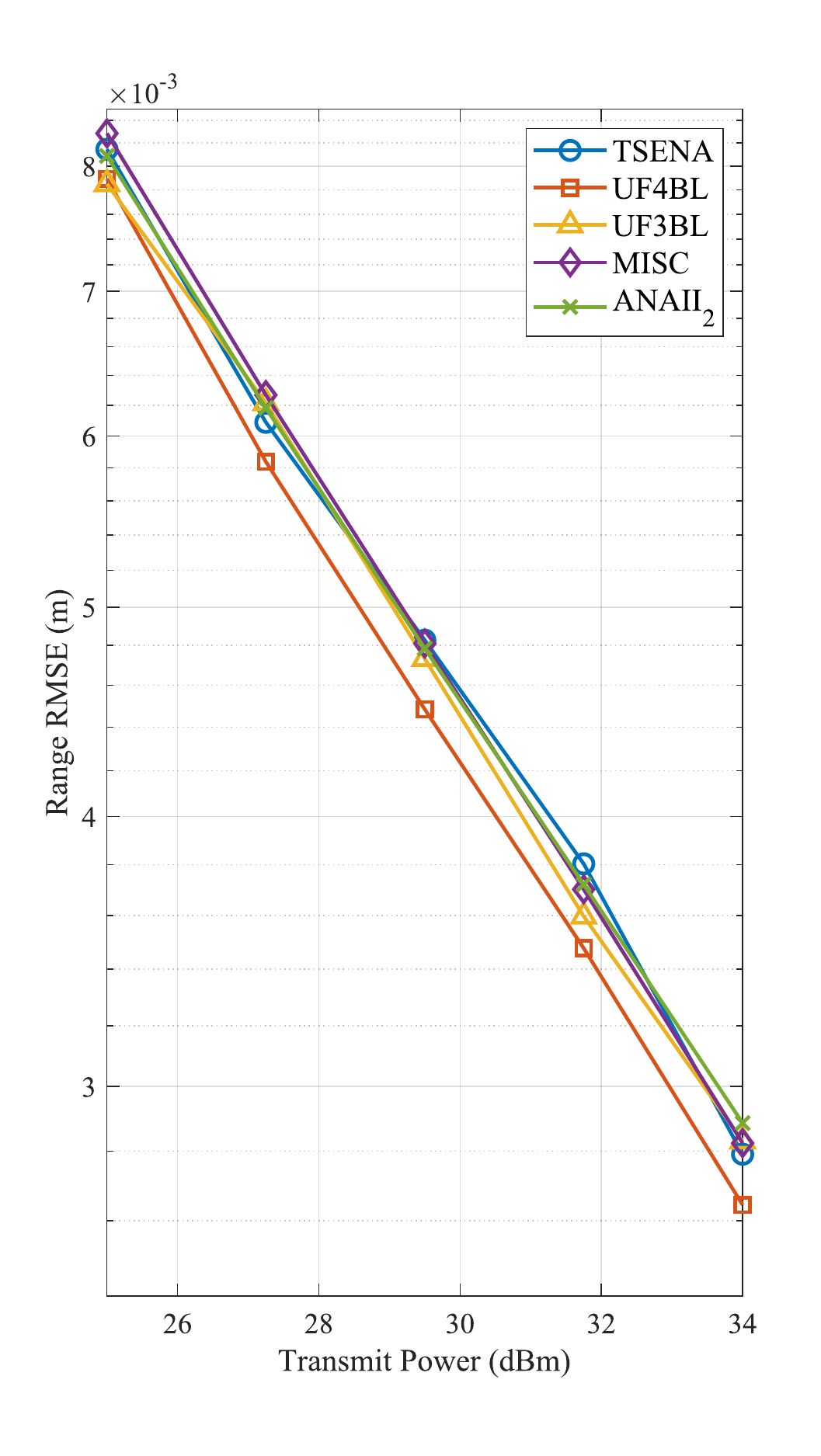}
        \caption{}
        \label{fig:sub1}
    \end{subfigure}
    \begin{subfigure}[b]{0.23\textwidth}
        \includegraphics[width=\linewidth]{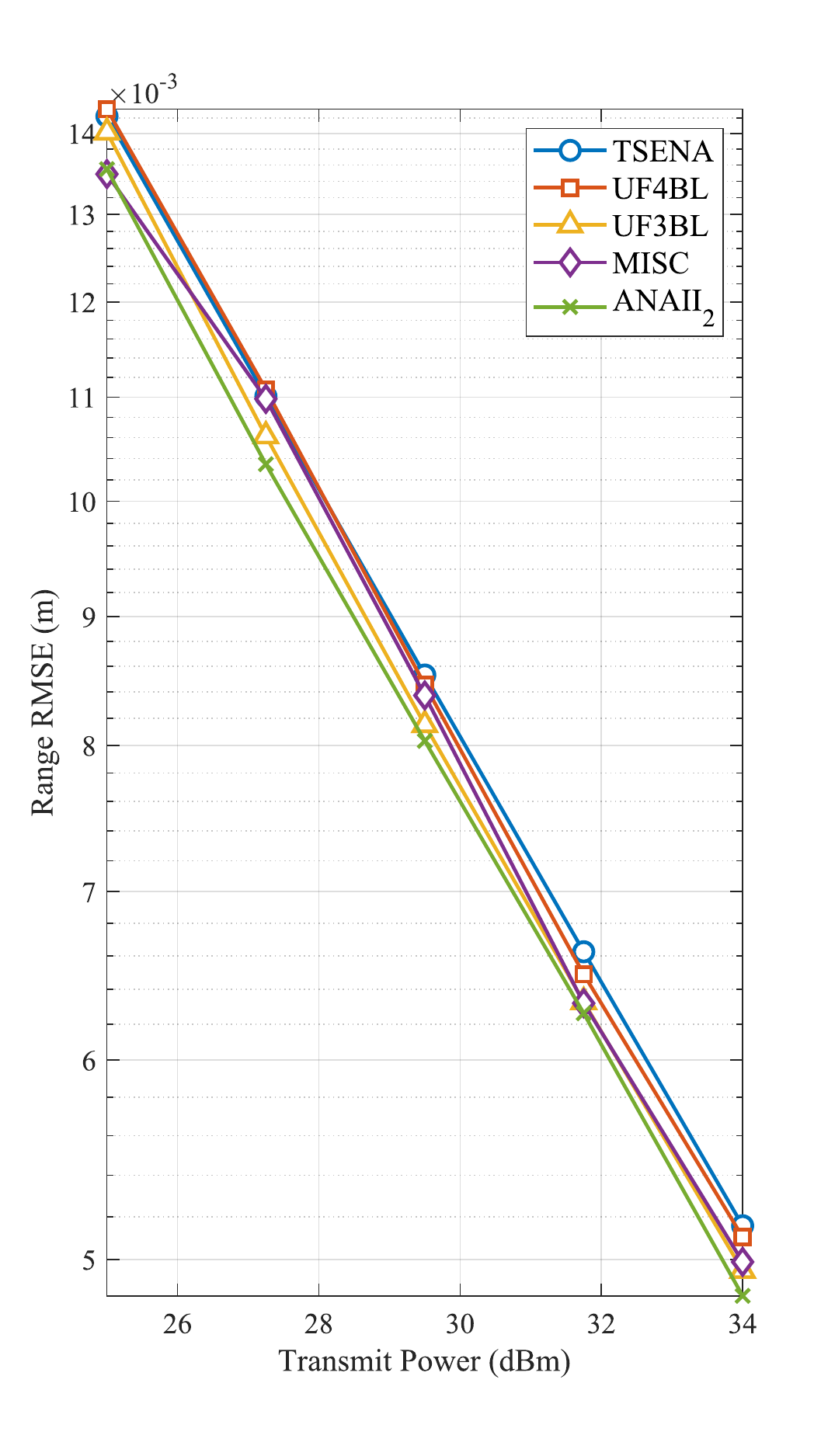}
        \caption{}
        \label{fig:sub2}
    \end{subfigure}
\caption{RMSE of range estimation versus transmit power with different CORSET\#0  Configuration: (a) 24RBs, (b) 48RBs}
    \label{fig:two_images}
\end{figure}

Fig. \ref{fig:two_images} illustrates the  performance of range estimation versus transmit power under different array layouts, where the Fig. \ref{fig:two_images}(a) and \ref{fig:two_images}(b) correspond to the 24RB and 48RB CORSET\#0 configurations, respectively. Since the frequency-domain configuration schemes primarily affect range estimation, only the ranging RMSE is evaluated. As shown in Fig. \ref{fig:two_images}(a), for the 24RB configuration, the UF4BL scheme achieves the best performance, which is attributed to its lower mutual coupling. Conversely, for the 48RB configuration in Fig. \ref{fig:two_images}(b), the ANAII-2 scheme provides the highest accuracy due to the increased uDoF it offers. Furthermore, we observe that for a given transmit power, the increased bandwidth from 24RBs to 48RBs does not improve sensing performance. This can be attributed to the decrease in the SNR per RE as the total power is distributed over a larger number of REs.

\subsection{Estimation Accuracy for Multiple Targets}

To evaluate the multi-target performance, we employ the optimal sub-pattern assignment (OSPA) metric\cite{pucci2022system}, which provides a unified performance assessment by jointly capturing both localization and cardinality errors (i.e., the difference between the true and estimated number of targets) in a single value.

Given the ground truth set of $L$ target positions $P = \{p_1, \dots, p_L\}$, and the estimated set of $\hat{L}$ positions $\hat{P} = \{\hat{p}_1, \dots, \hat{p}_{\hat{L}}\}$, the distance between any true position $p$ and estimated position $\hat{p}$ is defined as
\begin{equation}
    d(p, \hat{p}; \bar{c}) = \min\{\bar{c}, ||p - \hat{p}||_2\},
\end{equation}
where $\bar{c} > 0$ is the cutoff parameter that determines how the metric penalizes cardinality error with respect to the localization error. For the case  $L \le \hat{L}$, the OSPA metric of order $q$ is defined as \cite{schuhmacher2008consistent}
\begin{equation}
\begin{aligned}
    &\overline{d}_{q}(P, \hat{P}; \bar{c})\\ 
    &= \bigg( \frac{1}{\hat{L}} \bigg( \min_{\pi \in \Pi_{\hat{L}}} \sum_{l=1}^{L} & \left(d(p_{l}, \hat{p}_{\pi(l)}; \bar{c})\right)^{q} 
     + \bar{c}^{q}(\hat{L} - L) \bigg) \bigg)^{1/q},
\end{aligned}
\end{equation}
where $\Pi_{\hat{L}}$ is the set of permutations of $\{1, 2, \dots, \hat{L}\}$ and $\pi(l)$ is the index of the estimated point in $\hat{P}$ assigned to the $l$-th true point $p_l$ under a specific permutation $\pi\in \Pi_{\hat{L}}$. In particular, the metric is computed symmetrically as $\overline{d}_{q}(P, \hat{P}; \bar{c}) = \overline{d}_{q}(\hat{P}, P; \bar{c})$ if $L > \hat{L}$. Moreover, the order $q$ determines the sensitivity to outliers, while the cutoff $\bar{c}$ balances the localization and cardinality error penalties. In our simulations, we follow the recommendations in \cite{schuhmacher2008consistent} and set $q=2$ and $\bar{c}=10$~m to ensure a proper balance between these error components. Specifically, we consider $L=5$ targets, whose parameters are randomly drawn from uniform distributions: ranges in $[5, 85]$~m, velocities in $[-35, 35]$~m/s, and angles in $[-35^\circ, 35^\circ]$. We employ this setup to validate the performance of our proposed two-stage (coarse-to-fine) detection scheme in a multi-target environment.

\begin{figure}[h]
\centerline{\includegraphics[width=0.45\textwidth]{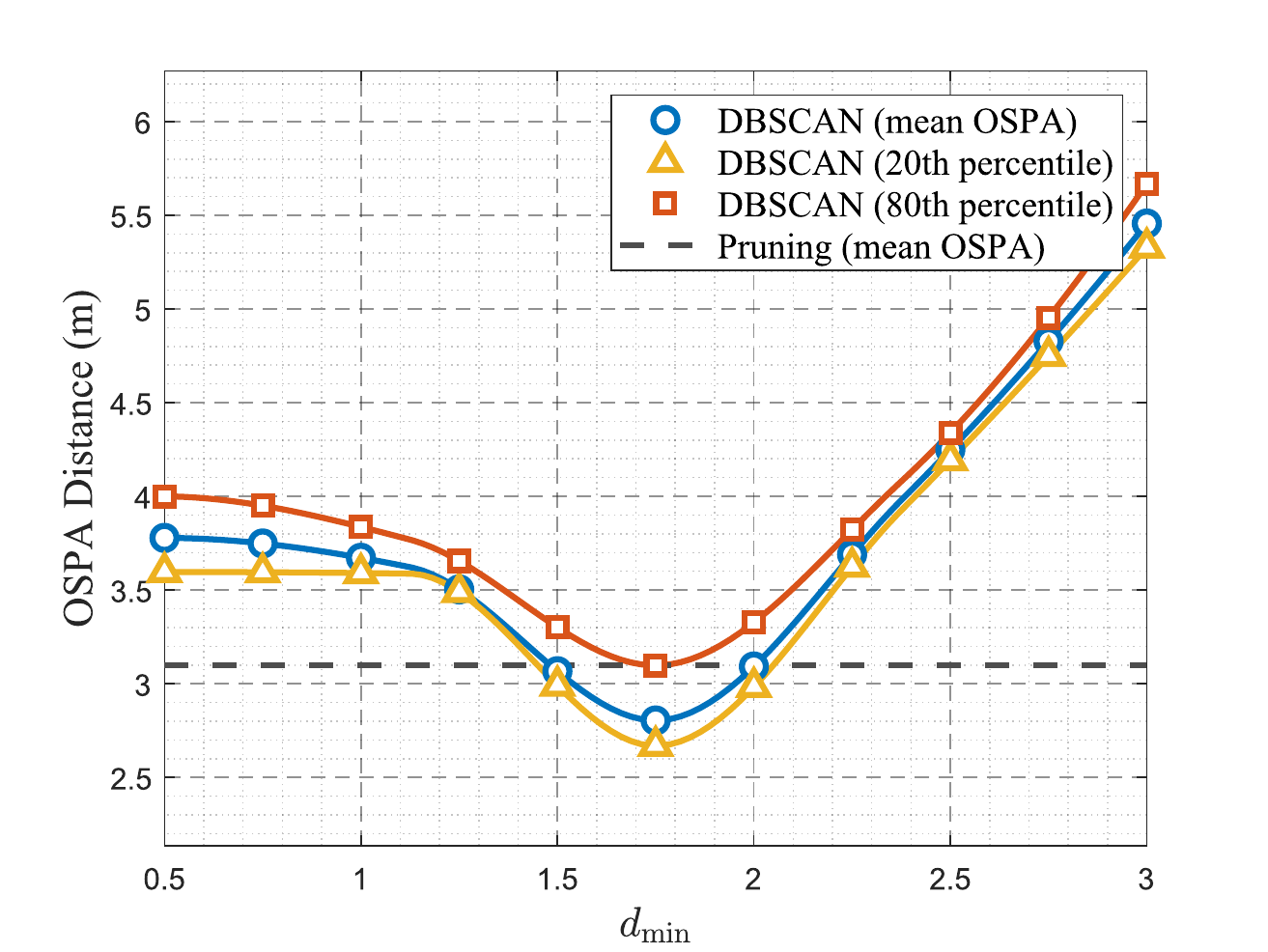}}
\caption{OSPA distance versus the minimum distance threshold $d_{\text{min}}$ of DBSCAN.}
\label{fig:OSPA_vs_dmin}
\end{figure}

Fig. \ref{fig:OSPA_vs_dmin} illustrates the OSPA performance of the proposed DBSCAN-based algorithm versus the minimum distance threshold $d_{\text{min}}$, compared against the Pruning algorithm from \cite{pucci2022system}. For the Pruning algorithm, parameters are set to $\epsilon_r = \Delta r$ and $\epsilon_v = 3 \Delta v$ as recommended in \cite{pucci2022system} to achieve its optimal performance, which is plotted as the black dashed line (mean OSPA). For our proposed algorithm, the mean, 20th, and 80th percentile curves are presented. 

The OSPA distance for the DBSCAN-based fusion algorithm exhibits a U-shaped trend, highlighting a critical trade-off. Specifically, when the threshold $d_{\text{min}}$ is too low, the algorithm  fail to associate detections from the same target (e.g., across different beams), resulting in an overestimation of the target count and increasing the cardinality error. Conversely, a large $d_{\text{min}}$ risks merging detections from distinct targets that are closely spaced in the range-Doppler-angle domain, which introduces both localization and cardinality errors. As shown,  the DBACAN-based fusion algorithm achieves an optimal balance between both error types around $d_{\text{min}} = 1.7$. With this proper hyperparameter selection, our method demonstrates superior performance over the benchmark.

\begin{figure}[h]
\centerline{\includegraphics[width=0.4\textwidth]{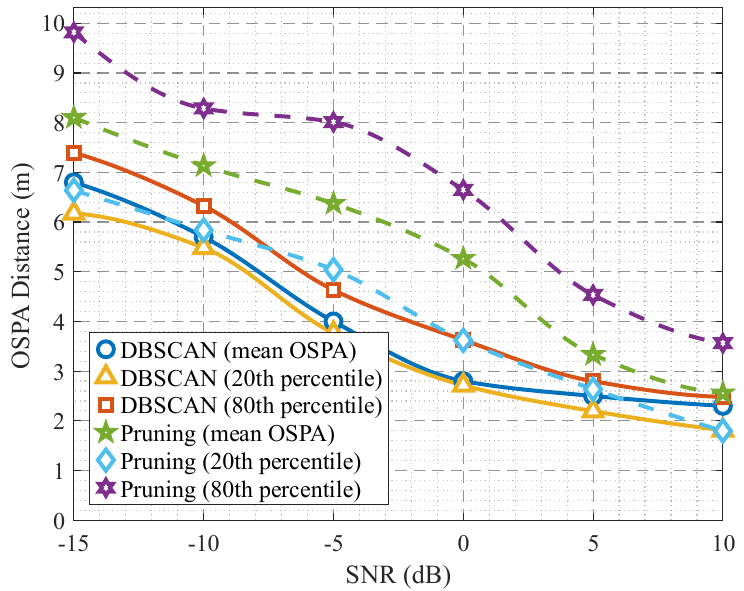}}
\caption{OSPA distance versus SNR.}
\label{fig:OSPA_vs_SNR}
\end{figure}

Fig. \ref{fig:OSPA_vs_SNR} depicts the OSPA distance versus SNR for the proposed DBSCAN-based fusion algorithm and the Pruning benchmark \cite{pucci2022system}. The proposed algorithm consistently achieves better accuracy and greater robustness across the entire SNR range. This improved robustness is reflected in the narrower spread between its 20th and 80th percentile curves around the mean. In contrast, the Pruning algorithm exhibits larger performance fluctuations, which can be attributed to its selection mechanism: it retains only the single target candidate with the maximum matching score.  The clustering nature of DBSCAN, however, inherently suppresses such variability by aggregating the regional detections, resulting in more stable and reliable performance.

\section{CONCLUSION}
This paper proposed a  two-stage ISAC framework designed to support the emerging LAE by leveraging 5G NR networks. Specifically, the framework addressed the fundamental sensing resolution limitations of standard broadcast signals. In Stage I, we first fused RDMs from the SSB, Type\#0-PDCCH, and SIB1 for robust initial detection. In Stage II, we employed the novel  SPS, inspired by sparse array theory and inserted into unused SIB1 resources, to overcome the resolution limitations.  To this end, a corresponding high-resolution algorithm based on the WUP estimator and the RELAX-based iterative method was developed to accurately extract off-grid range and velocity parameters from these sparse pilots.  Finally, the DBSCAN-based clustering scheme was employed  to prune redundant detections. Comprehensive simulation results demonstrated that the proposed framework achieves superior estimation accuracy at a significantly lower computational cost compared to conventional benchmark methods. Future work will focus on investigating the framework's performance in more complex, dense multi-target scenarios and exploring further optimization of the SPS patterns in 5G NR.

\begin{appendices}
\section{}

In this appendix, we derive the received frequency-domain signal model for a MIMO-OFDM system as presented in the main text. We begin with the continuous-time baseband signal models and proceed through the steps of sampling, CP-removal, and DFT.

Substituting \eqref{eq:sampled_signal_vector} into \eqref{eq:analog_signal}, and assuming a sufficiently long CP such that inter-symbol interference (ISI) is eliminated, the sampled signal from the $l$-th transmitted symbol is
\begin{align}
\mathbf{\hat{y}}(j,l) \approx \sum_{n=1}^{N_{\text{target}}} &\beta_n e^{j 2 \pi f_{D,n} t_{j,l}} \mathbf{b}_\text{r}(\theta_n) \mathbf{a}^H(\theta_n) \nonumber \\
&\times \sum_{i=0}^{N_{\mathrm{c}}-1} [\mathbf{X}]_{i,l} e^{j 2 \pi i \Delta f (t_{j,l} - \tau_n)},
\label{eq:app_discrete_time_signal}
\end{align}
where $t_{j,l} = lT + j/F_\text{s} + T_{\text{CP}}$ and the noise term is omitted for clarity.

An $N_s$-point Discrete Fourier Transform (DFT) is then applied to $\mathbf{\hat{y}}(j,l)$ along the index $j$ to obtain the frequency-domain signal $\mathbf{y}(k,l)$:
\begin{align}
\mathbf{y}(k,l) &= \sum_{j=0}^{N_s-1} \mathbf{\hat{y}}(j,l) e^{-j2\pi kj/N_s} \nonumber \\
&\approx \sum_{j=0}^{N_s-1} \sum_{n=1}^{N_{\text{target}}} \beta_n e^{j 2 \pi f_{D,n} t_{j,l}} \mathbf{b}_\text{r}(\theta_n) \mathbf{a}^H(\theta_n) \nonumber \\ 
& \quad \times \sum_{i=0}^{N_{\mathrm{c}}-1} [\mathbf{X}]_{i,l} e^{j 2 \pi i \Delta f (t_{j,l} - \tau_n)} e^{-j2\pi kj/N_s}.
\label{eq:app_dft_start}
\end{align}
By rearranging the summations and substituting the definition of $t_{j,l}$, \eqref{eq:app_dft_start} can be written as
\begin{align}
\mathbf{y}(k, l) \approx \sum_{n=1}^{N_{\text{target}}}& \beta_{n} \mathbf{b}_{\mathrm{r}}(\theta_{n}) \mathbf{a}^{H}(\theta_{n}) \sum_{i=0}^{N_{\mathrm{c}}-1}[\mathbf{X}]_{i, l} e^{-j 2 \pi i \Delta f \tau_{n}} \nonumber \\
&\times \left(\sum_{j=0}^{N_{s}-1} e^{j 2 \pi\left(f_{D, n}+i \Delta f\right) t_{j, l}} e^{-j 2 \pi k j / N_{s}}\right).
\label{eq:app_rearranged}
\end{align}
The summation over $j$ can be simplified. First, we expand $t_{j,l}$ and factor out the term that is constant with respect to $j$:
\begin{align}
    &\sum_{j=0}^{N_{s}-1} e^{j 2 \pi\left(f_{D, n}+i \Delta f\right) (lT+T_{CP}+j/F_s)} e^{-j 2 \pi k j / N_{s}} \nonumber \\
    &= e^{j 2 \pi\left(f_{D, n}+i \Delta f\right)\left(l T+T_{C P}\right)} \nonumber \\
    & \quad \times \sum_{j=0}^{N_{s}-1} e^{j 2 \pi\left(f_{D, n}+i \Delta f\right) (j/F_s)} e^{-j 2 \pi k j / N_{s}}.
\label{eq:app_factor_out}
\end{align}
Using the standard OFDM relation $F_s = N_s \Delta f$, the summation term in \eqref{eq:app_factor_out} becomes
\begin{equation}
\sum_{j=0}^{N_{s}-1} e^{j 2 \pi \frac{j}{N_{s}}\left(i-k+f_{D, n} / \Delta f\right)}.
\label{eq:app_j_sum}
\end{equation}
For scenarios where the Doppler shift is much smaller than the subcarrier spacing ($f_{D, n} \ll \Delta f$), we can approximate $f_{D, n}/\Delta f \approx 0$. This approximation neglects the inter-carrier interference (ICI). Under this condition, \eqref{eq:app_j_sum} simplifies to the  delta function
\begin{equation}
\sum_{j=0}^{N_{s}-1} e^{j 2 \pi \frac{j}{N_{s}}(i-k)}=N_{s} \cdot \delta_{i, k}.
\label{eq:app_kronecker_delta}
\end{equation}
Substituting \eqref{eq:app_kronecker_delta} back into \eqref{eq:app_rearranged} causes the summation over $i$ to collapse to a single term where $i=k$:
\begin{align}
\mathbf{y}(k, l) \approx \sum_{n=1}^{N_{\text{target}}}& \beta_{n} \mathbf{b}_{\mathrm{r}}(\theta_{n}) \mathbf{a}^{H}(\theta_{n}) [\mathbf{X}]_{k, l} e^{-j 2 \pi k \Delta f \tau_{n}} \nonumber \\
&\times e^{j 2 \pi\left(f_{D, n}+k \Delta f\right)\left(l T+T_{C P}\right)} \cdot N_s.
\label{eq:app_collapsed_sum}
\end{align}
After compensating for the known phase component $e^{j 2 \pi k \Delta f\left(l T+T_{C P}\right)}$ and neglecting the negligible term $e^{j 2 \pi f_{D, n}T_{C P}}$ and  the constant $N_s$,  \eqref{eq:app_collapsed_sum} simplifies to
\begin{equation}
\mathbf{y}(k,l) = \sum_{n=1}^{N_{\text{target}}} \beta_n \mathbf{a}_\text{t}^H(\theta_n) [\mathbf{X}]_{k,l} \, e^{j k \omega_r(R_n)} e^{j l \omega_v(v_n)} \mathbf{b}_\text{r}(\theta_n) + \mathbf{z}(k,l),
\label{eq:dft_signal_vector_final}
\end{equation}
where $\omega_r(R_n) \triangleq -4\pi \Delta f R_n/c$ and $\omega_v(v_n) \triangleq 2\pi f_{D,n} T$ respectively.

\section{}

\subsection{Stacking Order Convention}
To facilitate the joint estimation of all target parameters, we aggregate the observations from all receive antennas. First, for each receive antenna $p$, we form a data vector $\mathbf{b}_p = \text{vec} (\mathbf{Y}_p) \in \mathbb{C}^{MN \times 1}$ , where $[\mathbf{Y}_p]_{i,l}= [\mathbf{y}(i,l)]_p$. Then the total observation vector $\mathbf{b} \in \mathbb{C}^{N_rMN \times 1}$ is expressed as:
\begin{equation}
    \mathbf{b} = \begin{bmatrix}
        \mathbf{b}_0 \\ \mathbf{b}_1 \\ \vdots \\ \mathbf{b}_{N_r-1}
    \end{bmatrix}.
\end{equation}

\subsection{Contribution of the $n$-th Target}
For the $n$-th target, we define the time-frequency phase vector $\mathbf{s}_n \in \mathbb{C}^{M N \times 1}$ with elements arranged as
\begin{equation}
\mathbf{s}_n = \begin{bmatrix}
\mathbf{a}_\text{t}^H(\theta_n) [\mathbf{X}]_{0,0} e^{j0\cdot\omega_r(R_n)} e^{j0\cdot\omega_v(v_n)} \\
\mathbf{a}_\text{t}^H(\theta_n) [\mathbf{X}]_{1,0} e^{j1\cdot\omega_r(R_n)} e^{j0\cdot\omega_v(v_n)} \\
\vdots \\
\mathbf{a}_\text{t}^H(\theta_n) [\mathbf{X}]_{M-1,0} e^{j(M-1)\cdot\omega_r(R_n)} e^{j0\cdot\omega_v(v_n)} \\
\mathbf{a}_\text{t}^H(\theta_n) [\mathbf{X}]_{0,1} e^{j0\cdot\omega_r(R_n)} e^{j1\cdot\omega_v(v_n)} \\
\vdots \\
\mathbf{a}_\text{t}^H(\theta_n) [\mathbf{X}]_{M-1,N-1} e^{j(M-1)\cdot\omega_r(R_n)} e^{j(N-1)\cdot\omega_v(v_n)}
\end{bmatrix}.
\end{equation}
The contribution of the $n$-th target to the complete received vector is given by
\begin{equation}
\begin{aligned}
\mathbf{y}_n &= \beta_n \mathbf{b}_\text{r}(\theta_n) \otimes \mathbf{s}_n,\\
& =\beta_n \mathbf{b}_\text{r}(\theta_n) \otimes\left( \mathbf{T}^T \mathbf{a}_\text{t}^*(\theta_n) \odot \left( \mathbf{a}_v\left(v_n\right) \otimes \mathbf{a}_r\left(R_n\right) \right) \right) 
\end{aligned}
\end{equation}
where $\mathbf{a}_r(R_n) = [1, e^{j\omega_r(R_n)}, \ldots, e^{j(M-1)\omega_r(R_n)}]^T \in \mathbb{C}^{M \times 1}$, $\mathbf{a}_v(v_n) = [1, e^{j\omega_v(v_n)}, \ldots, e^{j(N-1)\omega_v(v_n)}]^T \in \mathbb{C}^{N \times 1}$ and $\mathbf{T} \in \mathbb{C}^{N_t \times M N}$ is formed by stacking $[\mathbf{X}]_{i,l}$ in column-major order.

\subsection{Complete Signal Model}
The full received signal is the superposition of all target contributions plus noise
\begin{equation}
\mathbf{y} = \sum_{n=1}^{N_{\text{target}}} \beta_n \left( \mathbf{b}_\text{r}(\theta_n) \otimes \left[ \left( \mathbf{T}^T \mathbf{a}_\text{t}^*(\theta_n) \right) \odot \left( \mathbf{a}_v(v_n) \otimes \mathbf{a}_r(R_n) \right) \right] \right) + \mathbf{z}.
\end{equation}
Denoting the joint steering vector for the $n$-th target as
\begin{equation}
\mathbf{a}_n = \mathbf{b}_\text{r}(\theta_n) \otimes \left[ \left( \mathbf{T}^T \mathbf{a}_\text{t}^*(\theta_n) \right) \odot \left( \mathbf{a}_v(v_n) \otimes \mathbf{a}_r(R_n) \right) \right],
\end{equation} 
the model can then be written in the matrix form 
\begin{equation}
\mathbf{y} = \mathbf{A} \boldsymbol{\beta} + \mathbf{z},
\end{equation}
where $\mathbf{A} = [\mathbf{a}_1,  \mathbf{a}_2,\ldots,\mathbf{a}_{N_{\text{target}}}]$ and $\boldsymbol{\beta} = [\beta_1, \beta_2, \ldots, \beta_{N_{\text{target}}}]^T$.

Substituting these results back, the final expression for the partial derivative of the joint steering vector with respect to the angle $\theta_n$ is:
\begin{align}
    \frac{\partial \mathbf{a}_n}{\partial \theta_n} &= (\mathbf{J}_{\text{br}} \mathbf{b}_\text{r}(\theta_n)) \otimes \mathbf{s}_n \nonumber \\
    & \quad - \mathbf{b}_\text{r}(\theta_n) \otimes \left[ \left( \mathbf{T}^T \mathbf{J}_{\text{at}} \mathbf{a}_\text{t}^*(\theta_n) \right) \odot \left( \mathbf{a}_v(v_n) \otimes \mathbf{a}_r(R_n) \right) \right].
\end{align}

\begin{equation}
    \frac{\partial \mathbf{a}_{n}}{\partial \omega_{r}\left(R_{n}\right)}=\mathbf{b}_{\mathrm{r}}\left(\theta_{n}\right) \otimes\left[\left(\mathbf{T}^{T} \mathbf{a}_{\mathrm{t}}^{*}\left(\theta_{n}\right)\right) \odot\left(\mathbf{a}_{v}\left(v_{n}\right) \otimes\left(\mathbf{J}_{r} \mathbf{a}_{r}\left(R_{n}\right)\right)\right)\right]
\end{equation}

\begin{equation}
    \frac{\partial \mathbf{a}_{n}}{\partial \omega_{v}\left(v_{n}\right)}=\mathbf{b}_{\mathrm{r}}\left(\theta_{n}\right) \otimes\left[\left(\mathbf{T}^{T} \mathbf{a}_{\mathrm{t}}^{*}\left(\theta_{n}\right)\right) \odot\left(\left(\mathbf{J}_{v} \mathbf{a}_{v}\left(v_{n}\right)\right) \otimes \mathbf{a}_{r}\left(R_{n}\right)\right)\right]
\end{equation}
where $\mathbf{J}_{\text{br}}$, $\mathbf{J}_{\text{at}}$, $\mathbf{J}_{r}$ and $\mathbf{J}_{v}$ are   diagonal matrixs. $[\mathbf{J}_{\text{br}}]_{i,i}=-j2\pi(i-1)d_r \cos \theta_n/\lambda_c $ and $[\mathbf{J}_{\text{at}}]_{i,i}=-j2\pi(i-1)d_t\cos \theta_n/\lambda_c $, $\mathbf{J}_{r}=j \cdot \operatorname{diag}(0,1,2, \ldots, M-1)$, $\mathbf{J}_{v}=j \cdot \operatorname{diag}(0,1,2, \ldots, N-1)$

Based on the Theorem 1 in \cite{hu2024joint}, the CRLB w.r.t $\boldsymbol{\Theta}=\left[\boldsymbol{\theta}^{T}, \boldsymbol{\omega}_r^{T}, \boldsymbol{\omega}_v^{T}\right]^{T}$ can be given as 
\begin{equation}
    \begin{aligned}
\mathrm{CRLB}(\boldsymbol{\Theta}) & =\frac{\sigma_{v}^{2}}{2}\left[\operatorname{Re}\left\{\tilde{\mathbf{B}}^{H}\left(\bar{\mathbf{A}}^{H} \mathbf{P}_{\mathbf{A}}^{\perp} \bar{\mathbf{A}}\right) \tilde{\mathbf{B}}\right\}\right]^{-1},
\end{aligned}
\end{equation}
where 
\begin{equation}
\begin{aligned}
    \bar{\mathbf{A}} = \bigg [  & \frac{\partial \mathbf{a}_{1}}{\partial \theta_{1}}, \frac{\partial \mathbf{a}_{2}}{\partial \theta_{2}}, \ldots, \frac{\partial \mathbf{a}_{N_{\text{target}}}}{\partial \theta_{N_{\text{target}}}}, \\
    & \frac{\partial \mathbf{a}_{1}}{\partial \omega_r(R_1)}, \frac{\partial \mathbf{a}_{2}}{\partial \omega_r(R_2)}, \ldots, \frac{\partial \mathbf{a}_{N_{\text{target}}}}{\partial \omega_r(R_{N_{\text{target}}})}, \\
    &  \frac{\partial \mathbf{a}_{1}}{\partial \omega_v(v_1)}, \frac{\partial \mathbf{a}_{2}}{\partial \omega_v(v_2)}, \ldots, \frac{\partial \mathbf{a}_{N_{\text{target}}}}{\partial \omega_v(v_{N_{\text{target}}})} \bigg]
\end{aligned}
\end{equation}
\begin{equation}
    \mathbf{P}_{\mathbf{A}}^{\perp}=\mathbf{I}-\mathbf{A}\left(\mathbf{A}^{H} \mathbf{A}\right)^{-1} \mathbf{A},
\end{equation},\begin{equation}\tilde{\mathbf{B}}=\mathbf{I}_{3} \otimes \operatorname{diag}(\boldsymbol{\beta})
\end{equation}

\end{appendices}

\bibliography{ref.bib}
\bibliographystyle{ieeetr}

\end{document}